%Paper: hep-th/9311062
%From: tseytlin@surya11.cern.ch (Arkady Tseytlin)
%Date: Wed, 10 Nov 93 16:18:57 +0100
%Date (revised): Wed, 10 Nov 93 16:46:11 +0100
%Date (revised): Sat, 20 Nov 93 14:40:56 +0100

%%%%%%%%%%%%%%%%%%%%%%%%%%%%%%%%%%%%%%%%%%%%%%%%%%%%%%%%%%%%%%%%
\input harvmac

%%%%%%%%%%%%%%%%%%%%%%%%%%%%%%%%%%%%%%%%%%%%%%%%%%%%%%%%%%%
\def\s {\sigma }
\def \unn {{\underline n}}
\def \unm {{\underline m}}
\def \unk {{\underline k}}
\def \unl {{\underline l}}
\def \bk  {{\overline k}}
\def \bl  {{\overline  l}}
\def \bm  {{\overline m}}
\def \bn  {{\overline n}}

\def \E {{\tilde E}}

\def \B { {\bar B}}
\def \L {\Lambda}
\def \Qab {Q_{ab} }
\def \M { {\cal M}}
\def \C {{\cal C}}
\def \sm {$\s$-model}
\def \sms {$\s$-models\ }
\def \bb  {{\bar \b }}

\def \dim  {{\rm dim \ }}
\def \ra {\rightarrow}
\def \bd {\bar \del}

 \def \k1 {{1\over
k}} \def \bh { {\bar h} } \def \ov { \over }
  \def \B { { \bar B }}

\def \a {\alpha}
\def \b {\beta}

\def \Tr {{\ \rm Tr \ }}

\def \det {{\ \rm det \ }}

\def \l {\lambda}
\def \1p {{1\over  \pi }}
\def \2p {{{1\over  2\pi }}}
\def \4p {{ {1\over 4 \pi }}}
\def \8p {{{1\over 8 \pi }}}
\def \P^* { P^{\dag } }
\def \p {\phi}

\def \M {{\cal M}}
\def \m {\mu }
\def \n {\nu}

\def\g {\gamma}
\def \r {\rho}
\def \k {\kappa }
\def \d {\delta}

\def \s {\sigma}

\def \fourth {{{1\over 4}}}

\def \e#1 {{{\rm e}^{#1}}}

\def \eq#1 {\eqno {(#1)}}
%%%%%%%%%%%%%%%%%%%%%%%%%%%%%%%%%%%%%%%%%%%%
\def \sm {$\s$-model\ }\def \B  {{ \tilde B }}
\def \I { { I }}

\def \bd  {{ \bar \del }}

\def \bd  { \bar \del }

\def \ov {\over }

\def \A  { {\bar A} }

\def \C {{\cal C }}
%%%%%%%%%%%%%%%%%%%%%%%%%%%%%%%%%%%%%%%
\def \H {{\cal H}}

\def \p {\phi}

\def \s {\sigma}

\def \r {\rho}
\def \d {\delta}
\def \l {\lambda}
\def \m {\mu}
\def \g {\gamma}
\def \n {\nu}

\def \e#1 {{{\rm e}^{#1}}}

\def \B {{\bar B}}
\def \H {{\cal H}}
\def \P {{ \Pi}}

\def \J {\bar J }
\def\np {  Nucl. Phys. }
\def \pl { Phys. Lett. }
\def \mpl { Mod. Phys. Lett. }
\def \prl { Phys. Rev. Lett. }
\def \pr  { Phys. Rev. }

\def \cmp { Commun. Math. Phys. }
\def \ijmp { Int. J. Mod. Phys. }
%%%%%%%%%%%%%%%%%%%%%%%%%%%%%%%%%%%%%%%%%
%%%%%%%%%%%%%%%%%%%%%%%%%%%%%%%%
\lref \nap { M. Henningson and C. Nappi,  \pr D48(1993)861.  }
\lref \Ve    { K.M. Meissner and G. Veneziano, \pl B267(1991)33;
A. Sen,  \pl  B271(1991)295.}

\lref \givkir {A. Giveon and E. Kiritsis,  preprint CERN-TH.6816/93,
RIP-149-93. }
\lref \coq {R. Coquereaux and A. Jadcyzk, {\it Riemannian Geometry, Fiber
Bundles, Kaluza--Klein
Theories  and all that...}(World Scientific, Singapore, 1988).}

 \lref  \rocver { M. Ro\v cek and E.
Verlinde, \np B373(1992)630.}
 \lref \kumar {A. Kumar,
preprint CERN-TH.6530/92.}
\lref  \hussen {  S. Hussan and A. Sen,  \np B405(1993)143. }
\lref \kiri {E. Kiritsis, \np B405(1993)109. }

\lref \givroc { A. Giveon and M. Ro\v cek, \np B380(1992)128. }

\lref \chaud {S. Chaudhuri and J.A. Schwartz, \pl B219(1989)291.}

\lref \tsdu { A.A. Tseytlin, \pl B242(1990)163; \np B350(1991)395.  }

\lref \sto {M. Stone, preprint NSF-ITP-89-97;   K. Harada,  \ijmp A6(1991)3399.
}

\lref \kart { D. Karabali, Q-H. Park and  H.J. Schnitzer, \np B323(1989)572;
\pl B205(1988)267.
 }
\lref \thirpot {D. Karabali, H.J. Schnitzer  and K. Tsokos, \np B294(1987)412;
I. Antoniadis, C. Bachas and C. Kounnas, \pl B200(1988)297.}
\lref \boul {L.S. Brown and R.I. Nepomechie, \pr D35(1987)3235.  }
\lref \dash {R. Dashen and Y. Frishman, \pr D11(1975)2781.}
\lref \kut {D. Kutasov, \pl B227(1989)68; \pl B233(1989)369.}
\lref \gmm {E. Guadagnini, M. Martellini and M. Mintchev, \pl B194(1987)69.}
\lref \gat  { D. A. Depireux, S.J. Gates and Q-H. Park, \pl B224(1989)364;
 S.J. Gates and W. Siegel, \pl B206(1988)631.
}
\lref \solo {O.A. Soloviev,  preprint QMW-93-19; hep-th/9307163.}

%S.J. Gates and O.A. Soloviev, \pl B294(1992)342;
 % O.A. Soloviev, \mpl A8(1993)301;

\lref\wi {E. Witten,  Commun. Math. Phys. 144(1992)189.}
\lref \nov {S.P. Novikov,  Sov. Math. Dokl. 37(1982)3. }
\lref \fuch { J. Fuchs, \np B286(1987)455 and  B318(1989)631. }
\lref \sus { R. Rohm,  \pr D32(1985)2845.}
\lref \suss {   H.W. Braden, \pr D33(1986)2411.}
\lref   \red { A.N. Redlich and H.J. Schnitzer, \pl B167(1986)315 and
B193(1987)536(E);  A. Ceresole,
 A. Lerda, P. Pizzochecco
 and
P. van Nieuwenhuizen, \pl
 B189(1987)34.}
 \lref \div  { P. Di Vecchia,  V. Knizhnik, J. Peterson and P. Rossi, \np
B253(1985)701.}
\lref \thir { C. Destri, \pl B210(1988)173; B213(1988)565(E);
 A. Bondi, G. Curci, G. Paffuti and P. Rossi, \pl B216(1989)345. }

\lref \nem {D. Nemeschansky and S. Yankielowicz, \prl 54(1985)620; 54(1985)1736
(E).}

\lref \mor {A.Yu. Morozov, A.M. Perelomov, A.A. Rosly, M.A. Shifman and A.V.
Turbiner, \ijmp
A5(1990)803.}

\lref \hal { M.B. Halpern and E.B. Kiritsis, \mpl A4(1989)1373; A4(1989)1797
(E).}

\lref \haly {M.B. Halpern and   J.P. Yamron,  Nucl.Phys.B332(1990)411.
}
\lref \hally {M.B. Halpern and   J.P. Yamron,
Nucl.Phys.
B351(1991)333.}

\lref \halpo{ M.B. Halpern, E.B. Kiritsis, N.A. Obers, M. Porrati and J.P.
Yamron,
\ijmp A5(1990)2275;  A. Giveon, M.B. Halpern, E.B. Kiritsis and  N.A. Obers,
\np B357(1991)655.}

\lref \halpoo{ M.B. Halpern  and  N.A. Obers,  \np B345(1990)607.}

\lref \morshi {  A.Yu. Morozov,  M.A. Shifman and A.V. Turbiner, \ijmp
A5(1990)2953.}

\lref \halpgiv {A. Giveon, M.B. Halpern, E.B. Kiritsis and  N.A. Obers,
\np B357(1991)655.}

\lref \halpe {M.B. Halpern, in: {\it Strings and Symmetries,} Proc. of Stony
Brook conference,
1991, p.447 (World Scientific,1992).}

\lref \halp { M.B. Halpern, E.B. Kiritsis, N.A. Obers, M. Porrati and J.P.
Yamron,
\ijmp A5(1990)2275;
  A.Yu. Morozov,  M.A. Shifman and A.V. Turbiner, \ijmp
A5(1990)2953;
A. Giveon, M.B. Halpern, E.B. Kiritsis and  N.A. Obers,
\np B357(1991)655.}
\lref \bpz {A.A. Belavin, A.M. Polyakov and A.B. Zamolodchikov, \np
B241(1984)333. }
\lref \gwz {      K. Bardakci, E. Rabinovici and
B. S\"aring, \np B299(1988)157;
 K. Gawedzki and A. Kupiainen, \pl B215(1988)119;
\np B320(1989)625. }

\lref \sen {A. Sen, preprint TIFR-TH-92-57. }

\lref \bcr {K. Bardakci, M. Crescimanno and E. Rabinovici, \np
B344(1990)344. }
\lref \Jack {I. Jack, D.R.T.  Jones and J. Panvel,  \np B393(1993)95. }
\lref \zam  { Al. B. Zamolodchikov, preprint ITEP 87-89. }

\lref \hor {J. Horne and G. Horowitz, \np B368(1992)444. }
\lref \tse { A.A. Tseytlin, \pl B264(1991)311. }
\lref \GWZNW  { P. Di Vecchia and P. Rossi, \pl  B140(1984)344;
 P. Di Vecchia, B. Durhuus  and J. Petersen, \pl  B144(1984)245.}
\lref \oal { O. Alvarez, \np B238(1984)61. }

\lref \ishi { N. Ishibashi, M.  Li and A. Steif, \prl 67(1991)3336. }
\lref  \kuma  { M. Ro\v cek and E. Verlinde, \np B373(1992)630; A. Kumar,
preprint CERN-TH.6530/92;
 S. Hussan and A. Sen,  preprint  TIFR-TH-92-61;  D. Gershon,
preprint TAUP-2005-92; X. de la Ossa and F. Quevedo, preprint NEIP92-004; E.
Kiritsis, preprint LPTENS-92-29. }

\lref \rocver { A. Giveon and M. Ro\v cek, \np B380(1992)128. }
\lref \frts {E.S. Fradkin and A.A. Tseytlin, \np B261(1985)1. }
\lref \mplt {A.A. Tseytlin, \mpl A6(1991)1721. }
\lref\bn {I. Bars and D. Nemeschansky, \np B348(1991)89.}
\lref \shif { M.A. Shifman, \np B352(1991)87.}
\lref\wittt { E. Witten, \cmp 121(1989)351; G. Moore and N. Seiberg, \pl
B220(1989)422.} \lref \chernsim { E. Guadagnini, M. Martellini and M.
Mintchev, \np B330(1990)575;
L. Alvarez-Gaume, J. Labastida and A. Ramallo, \np B354(1990)103;
G. Giavarini, C.P. Martin and F. Ruiz Ruiz, \np B381(1992)222; preprint
LPTHE-92-42.}
\lref \shifley { H. Leutwyler and M.A. Shifman, \ijmp A7(1992)795. }
\lref \polwig { A.M. Polyakov and P.B. Wiegman, \pl B131(1984)121; \pl
B141(1984)223.  }
\lref \polles { A. Polyakov, in: {\it Fields, Strings and Critical Phenomena},
  Proc. of Les Houches 1988,  eds.  E. Brezin and J. Zinn-Justin
(North-Holland,1990).   }
\lref \kutas {
D. Kutasov, \pl B233(1989)369.} \lref \karabali { D. Karabali, Q-H. Park, H.J.
Schnitzer and
Z. Yang, \pl B216(1989)307;  D. Karabali and H.J. Schnitzer, \np B329(1990)649.
}
\lref \ginq {P. Ginsparg and F. Quevedo,  \np B385(1992)527. }
\lref \bah {K. Bardakci and M.B. Halpern, \pr D3(1971)2493;
 M.B. Halpern, \pr D4(1971)2398.}
\lref \gko  {   P. Goddard,
A. Kent and D. Olive, \pl B152(1985)88; \cmp 103(1986)303.   }
\lref \dvv  { R. Dijkgraaf, H. Verlinde and E. Verlinde, \np B371(1992)269. }
\lref \kniz {  V. Knizhnik and A. Zamolodchikov, \np B247(1984)83. }

\lref \witt { E. Witten, \cmp 92(1984)455.}
\lref \wit { E. Witten, \pr D44(1991)314.}
\lref \anton { I. Antoniadis, C. Bachas, J. Ellis and D.V. Nanopoulos, \pl B211
(1988)393.}
\lref \bsfet {I. Bars and  K. Sfetsos, \pr D46(1992)4510.}
\lref\bsglobal{I. Bars and  K. Sfetsos, \pr D46(1992)4495.}
%
% \pl B301(1993)183;

\lref \ts  {A.A. Tseytlin, \pl B268(1991)175. }
\lref \shts {A.S. Schwarz and A.A. Tseytlin, \np B399(1993)691. }

\lref\bsft { I. Bars, preprint USC-91-HEP-B3. }
\lref\BSthree{I. Bars and K. Sfetsos, Mod. Phys. Lett. { A7}(1992)1091.}
\lref\BShet{I. Bars and K. Sfetsos, Phys. Lett. {B277}(1992)269.}
\lref\bs { I. Bars and K. Sfetsos,  Phys. Rev. {D48}(1993)844.}
\lref\bb  { I. Bars, \np B334(1990)125. }
\lref \tsw  { A.A. Tseytlin,\np B399(1993)601.}
\lref \ger { A. Gerasimov, A. Morozov, M. Olshanetsky, A. Marshakov and S.
Shatashvili, \ijmp
A5(1990)2495. }
\lref \hull {C.M. Hull,  \pl B206(1988)234; B212(1988)437. }

\lref \kir {{E. Kiritsis, \mpl A6(1991)2871.} }
\lref \nem {D. Nemeschansky and S. Yankielowicz, Phys.Rev.Lett. 54(1985)620;
54(1985)1736(E).}

\lref \br {A. Barut and R. Raczka, ``Theory of Group Representations and
Applications", p.120
 (PWN, Warszawa 1980). }
\lref \aat { A.A. Tseytlin, preprint CERN-TH.6804/93, hep-th/9302083.}
\lref \at {A.A. Tseytlin, preprint CERN-TH.6820/93. }
\lref \aps { S. De Alwis, J. Polchinski and R. Schimmrigk, \pl B218(1989)449. }
\lref \gny  { M.D. McGuigan, C.R. Nappi and S.A. Yost, \np B375(1992)421. }
\lref \tv {   A.A. Tseytlin and C. Vafa, \np B372(1992)443.}
\lref \ell {U. Ellwanger, J. Fuchs and M.G. Schmidt, \pl B203(1988)244; \np
B314(1989)175. }
\lref \ket { S.V. Ketov and O.A. Soloviev, \pl B232(1989)75; \ijmp
A6(1991)2971. }
\lref \GR {A. Giveon and M. Ro\v{c}ek, Nucl. Phys. B380(1992)128.}
\lref \giv {  A. Giveon, Mod. Phys. Lett. A6(1991)2843.}
\lref \GRV {A. Giveon, E. Rabinovici and G. Veneziano, Nucl. Phys.
B322(1989)167;
A. Shapere and F. Wilczek, \np B320(1989)669.}
\lref \GMR {A. Giveon, N. Malkin and E. Rabinovici, Phys. Lett. B238(1990)57.}
\lref \GS  {  A. Giveon and D.-J. Smit. Nucl. Phys.  B349(1991)168.}
\lref  \GP   {  A. Giveon and A. Pasquinucci, Phys. Lett. B294(1992)162.}
\lref \GG  { I.C. Halliday, E. Rabinovici, A. Schwimmer and M. Chanowitz, \np
B268(1986)413. }
%\lref \Ve    { K.M. Meissner and G. Veneziano, \pl B267(1991)33;
%M. Gasperini, J. Maharana and G. Veneziano, \pl B272(1991)277;
%A. Sen,  \pl  B271(1991)295.}
%%%%%%%%%%%%%%%%%%%%%%%
\lref \tye { S. Chung and S.-H. Tye, \pr D47(1993)4546.}
\lref  \GRT   { A. Giveon, E. Rabinovici and A.A. Tseytlin,  preprint
CERN-TH.6872/93,
RI-150-93, hep-th/9304155.}
 \lref \kz {  V.G. Knizhnik and A.B. Zamolodchikov, \np B247(1984)83. }
\lref \sfet { K. Sfetsos, preprint USC-93/HEP-S1, hep-th/9305074.}
\lref  \horne { J.H. Horne and G.T. Horowitz, \np B368(1992)444.}

\lref \givkir {A. Giveon and E. Kiritsis,  preprint CERN-TH.6816/93,
RIP-149-93. }
\lref \kumar { S.K. Kar and A. Kumar, \pl B291(1992)246;
 S. Mahapatra, \mpl A7(1992)2999;
S.K. Kar, S.P. Khastrig and G. Sengupta, \pr D47(1993)3643.
}
\lref \bow  {P. Bowcock, \np B316(1989)80. }
\lref \witten { E. Witten,  Commun. Math. Phys. 144(1992)189. }
\lref \sfts {K. Sfetsos and A.A. Tseytlin, preprint CERN-TH.69.. ,  to appear.}
\lref \gin  { P. Ginsparg and F. Quevedo, \np B385(1992)527.}
\lref \sft    {K. Sfetsos, \np B389(1993)424.}
\lref \who {
M. Crescimanno, \mpl A7(1992)489;
 I. Bars and K. Sfetsos, \mpl A7(1992)1091; \pl B277 (1992) 269;
  E.S. Fradkin and V.Ya. Linetsky, \pl B277(1992)73;
 A.H. Chamseddine, \pl B275(1992)63.
}

\lref\hor{
P. Horava, \pl { B278}(1992)101.}
\lref \rai {E. Raiten, ``Perturbations of a Stringy Black Hole'', Fermilab-Pub
91-338-T.
 }

\lref\cres{M. Crescimanno, \mpl A7(1992)489.}
\lref \frlin { E.S. Fradkin and V.Ya. Linetsky, \pl B277(1992)73\hb
 A.H. Chamseddine, \pl B275(1992)63. }

\lref \napwi  { C. Nappi and E. Witten, Phys. Lett. {B293}(1992)309. }
\lref \ST {K. Sfetsos and A.A. Tseytlin, preprint CERN-TH.6962/93,
hep-th/9308018.}
\lref \STT {K. Sfetsos and A.A. Tseytlin, preprint CERN-TH.6969/93,
hep-th/9310159.}

\lref \curci { G. Curci and G. Paffuti, \np B286(1987)399.   }
\lref \ajj {R.W. Allen, I. Jack and D.R.T. Jones, Z. Phys. C41(1988)323. }
\lref \ttt {A.A. Tseytlin, \pl B178(1986)349.}
\lref \hult {C.M. Hull and P.K. Townsend, \np B301(1988)197.}
\lref \jaj {I. Jack and D.R.T. Jones,  \pl B200(1988)453; \np B303(1988)260.}
\lref \tsey {A.A. Tseytlin. \pl B176(1986)92; \np B276(1986)391.}
 \lref \grwi {  D. Gross and E. Witten, \np B277(1986)1. }
\lref\BSslsu{I. Bars and K. Sfetsos, \pl  B301(1993)183.}
\lref\POLWIG { A.M. Polyakov and P.B. Wiegman, \pl  B141(1984)223.  }
\lref \schouten {J.A. Schouten, {\it  Ricci Calculus} (Springer, Berlin, 1954).
}
\lref \busch {T.H. Buscher, \pl B201(1988)466.}
\lref \nem {D. Nemeschansky and S. Yankielowicz, Phys.Rev.Lett. 54(1985)620;
54(1985)1736(E).}
\lref \efr {S. Elitzur, A. Forge and E. Rabinovici, \np B359 (1991)
581;
 G. Mandal, A. Sengupta and S. Wadia, Mod. Phys. Lett. A6(1991)1685. }

\lref \bcr {K. Bardakci, M. Crescimanno and E. Rabinovici, \np
B344(1990)344. }
\lref \Jack {I. Jack, D.R.T.  Jones and J. Panvel,  \np B393(1993)95. }
\lref \tse { A.A. Tseytlin, \pl B264(1991)311. }

\lref \dvv  { R. Dijkgraaf, H. Verlinde and E. Verlinde, \np B371(1992)269. }
\lref \kniz {  V. Knizhnik and A. Zamolodchikov, \np B247(1984)83. }

\lref \witt { E. Witten, \cmp 92(1984)455.}
\lref \wit { E. Witten, \pr D44(1991)314.}
\lref \anton { I. Antoniadis, C. Bachas, J. Ellis and D.V. Nanopoulos, \pl B211
(1988)393.}
\lref \ts  {A.A. Tseytlin, \pl B268(1991)175. }
\lref \kir {{E. Kiritsis, \mpl A6(1991)2871.} }

\lref \shts {A.S. Schwarz and A.A. Tseytlin,  preprint Imperial/TP/92-93/01
(1992). }
\lref \bush { T.H. Buscher, \pl B201(1988)466.  }

\lref \plwave { D. Amati and C. Klim\v cik, \pl B219(1989)443; G. Horowitz and
A. Steif, \prl 64(1990)260.}
\lref \faj { L.D. Faddeev and R. Jackiw, \prl 60(1988)1692.}

\lref \floja {R. Floreanini and R. Jackiw, \prl 59(1987)1873.}
\lref \chirwzw { Y. Frischman and J. Sonnenschein, \np B301(1988)346;
L. Mezinchescu and R. Nepomechie, \pr D37(1988)3067.}
\lref \son {J. Sonnenschein, \np B309(1988)752;   F. Bastianelli and P.van
Nieuwenhuizen,
\pl B217(1989)98; P. Salomonson, B.-S. Skagerstam  and A. Stern,  \prl
B62(1989)1817; S. Bellucci, M.
Golterman and D. Petcher, \np B326(1989)307; E. Abdalla and M. Abdalla, \pr
D40(1989)491;
M. Stone, \np B327(1989)339.}
 \lref \sigel { W. Siegel, \np B238(1984)307. }

\lref\bsft { I. Bars, preprint USC-91-HEP-B3. }

\lref\bs { I. Bars and K. Sfetsos,  \pr D48(1993)844. }
\lref \tsw  { A.A. Tseytlin,\np B399(1993)601.}
%%%%%%%%%%%%%%%%%%%%%%%%%%%%%%%%%%%%%%%%%%%%%%%%%%%%%%
\lref \call {C.G. Callan, D. Friedan, E. Martinec and M.J. Perry,
Nucl. Phys.B262(1985)593.}
\lref \witten { E. Witten,  Commun. Math. Phys. 144(1992)189. }
\lref \ajj {R.W. Allen, I. Jack and D.R.T. Jones, Z. Phys. C41(1988)323. }
%%%%%%%%%%%%%%%%%%%%%%%%%%%%%%%%%%%%%%%%%%%%%%%%%%%%%%%%%%%%%%%%%%%%
\lref \myers { R. Myers, Phys.Lett. B199(1987)371;
I. Antoniadis, C. Bachas, J. Ellis and D. Nanopoulos,
\pl B211(1988)393. }
\lref \parall { T.L. Curtright  and C.K. Zachos, \prl 53(1984)1799;
S. Mukhi, \pl B162(1985)345; S. de Alwis, \pl B164(1985)67. }
\lref \tsmpl {A.A. Tseytlin, \mpl A6(1991)1721.}

\lref \calg { C.G. Callan and Z. Gan, \np B272(1986)647. }
\lref \tss {A.A. Tseytlin, \pl B178(1986)34. }
\lref \fri  {D.H. Friedan, \prl 45(1980)1057; Ann. Phys. (NY) 163(1985)318. }
\lref \tsss { A.A. Tseytlin, \ijmp A4(1989)1257. }
\lref \osb { H. Osborn, Ann. Phys. (NY) 200(1990)1.}
\lref \jackk {I. Jack, D.R.T. Jones and D.A. Ross, \np B307(1988)130.}
\lref \jackkk {I. Jack, D.R.T. Jones and N. Mohammedi, \np B332(1990)333.}
\lref \shts  {A.S. Schwarz and A.A. Tseytlin, \np B399(1993)691.}

\lref \tseyt{A.A. Tseytlin,  preprint CERN-TH.6970/93, hep-th/9308042
(revised).}
\lref \giki{A. Giveon and E. Kiritsis, preprint CERN-TH.6816/93, RI-149-93,
hep-th/9303016 (revised).}

%%%%%%%%%%%%%%%%%%%%%%%%%%%%%%%%%%%%%%%%%%%%%%%%%%%%%%%%%%%%%%%
%%%%%%%%%%%%%%%%%%%%%%%%%%%%%%
\baselineskip8pt
\Title{\vbox
{\baselineskip8pt{\hbox{CERN-TH.7068/93}}{\hbox{hep-th/9311062}} }}
{\vbox{\centerline { On a  `universal' class of  WZW-type }\vskip2pt
 \centerline{  conformal   models }
}}
\vskip  -20 true pt
\centerline{A.A. Tseytlin\footnote{$^{*}$}{\baselineskip8pt
On leave  from Lebedev Physics
Institute, Moscow, Russia.
e-mail: tseytlin@surya3.cern.ch and tseytlin@ic.ac.uk} }
\smallskip
\centerline {\it Theory Division, CERN}
\centerline {\it
CH-1211 Geneva 23, Switzerland}
\smallskip
\centerline  {and}
\smallskip
\centerline {\it Theoretical Physics Group, Blackett Laboratory}
\centerline {\it Imperial College}
\centerline {\it London SW7 2BZ, U.K.}
\vskip 10 pt
\centerline {\bf Abstract}
\smallskip
\baselineskip8pt
\noindent

We consider a class of sigma models that  appears from
a generalisation of the gauged WZW model parametrised
by a constant matrix $Q$. Particular values of $Q$ correspond
to the standard gauged WZW models, chiral gauged WZW models
and a bosonised  version of the  non-abelian Thirring model.
The condition of conformal invariance of the models (to one loop
or $1/k$-order but exactly in $Q$) is derived and is represented
as an algebraic equation on $Q$. Solving this equation we
demonstrate explicitly the conformal invariance of the sigma models
associated with  arbitrary $G/H$ gauged and chiral gauged WZW
theories as well as of the models that can be represented as
WZW model perturbed by integrably marginal operators (constructed
from currents of the Cartan subalgebra $H_c$ of $G$). The latter
models can be also interpreted as $[G\times H]/H$ gauged WZW models
and have the corresponding target space couplings (metric,
antisymmetric tensor and dilaton) depending on an arbitrary constant
matrix which parametrises an embedding of the abelian subgroup $H$
(isomorphic to $H_c$) into $G\times H$. We discuss the relation of
our conformal invariance equation to the large $k$ form of the master
equation of the affine-Virasoro construction. Our equation describes
`reducible' versions of some `irreducible' solutions (cosets) of the
master equation. We suggest a  classically non-Lorentz-invariant sigma
models that may correspond to other solutions of the master equation.

\vskip 20 pt
\noindent
{CERN-TH.7068/93}
\smallskip
\noindent
 {November  1993}
\Date { }
%\draftmode
\noblackbox
\baselineskip 20pt plus 2pt minus 2pt

%%%%%%%%%%%%%%%%%%%%%%%%%%%%%%%%%%%%%%%
%%%%%%%%%%%%%%%%%%%%%%%%%%%%%%%%
\newsec {Introduction}

There  exists a large class of solutions  of string equations  related to
(gauged) WZW models that  were actively discussed recently.
  This suggests  to look for  other  conformal solutions based on similar
WZW-type models as well as to try to understand the  existence of
 known solutions  in a systematic way.
 Below we shall consider a `universal' model that
 contains  other  known  models   as special cases.
Particular cases   include    \sms which correspond to gauged  and chiral
gauged WZW theories,
bosonised version \kart\ of the non-abelian  Thirring  model \dash\  and  WZW
model perturbed by
integrably marginal  $J\J$   operators  \chaud.

Part of the original motivation behind this work was to   explore a possibility
to derive the
Virasoro master equation \hal\mor\ as a condition of conformal invariance of a
standard
off shell Lorentz invariant  sigma model. Since the
affine-Virasoro construction  \hal\mor (see also
\hally\halpo\morshi\halpoo\halpe)
contains  the
 affine-Sugawara  and coset models \bah\gko\ as special cases  and is
parametrised by a constant
matrix $L^{ab}$  it is
natural to start with a \sm  that also generalises the corresponding field
theories -- the
WZW model \witt\nov\ and the gauged WZW model \gwz\karabali\wi.\foot{The  $2d$
field theory action
suggested in \hally\ does not include the case of the coset models and is not
of the  standard \sm
type since it   contains some extra  fields (Lagrange multipliers) which are
necessary for its
Lorentz invariance.}

In Section 2 we shall present the classical Lagrangian and Hamiltonian of our
model  which depends
on a constant matrix $Q$  and  explain  how   various  known models correspond
to  special
values  of
$Q$.

In Section 3 we shall put  the action  in the form of a \sm  one  and compute
the
corresponding conformal anomaly coefficients (`$\bar \b$-functions') in the
leading order
approximation in $\a'=2/k$.  We shall find  that the conditions of conformal
invariance  reduce to an
algebraic equation  for   the basic  matrix $K$ (related to $Q$) which is equal
to the
constant part of the target space metric in normal coordinate system.
  This equation  can be derived from a `central charge action'.

In Section 4 we shall study the solutions of the conformal invariance  equation
and, in particular,
demonstrate that the gauged {and} chiral gauged $G/H$   WZW \sms are conformal
invariant for
{\it arbitrary} simple $G$ and $H$. We shall also show  that the \sm
perturbation theory gives the
expected expressions for the central charges. We shall find another class of
solutions
with $K$ depending on an arbitrary constant $r\times r$ ($r={\rm rank\ } G$)
matrix $\r$
and interpret them in terms of $[G\times H]/H$ gauged WZW models with $H$ being
an abelian  group of
dimension $r$ and $\r$ being related to the coefficients that parametrise  an
embedding of $H$ into
$G\times H$.   The equation can be solved explicitly in the case of $G=SU(2)$
when no new
solutions (except the already mentioned above) are found.

In Section 5 we shall discuss the relation of our conformal invariance equation
to the large $k$ limit of the Virasoro master equation \hal\mor.   In general,
the  solutions of the
two equations form  different but intersecting sets (with the  common solutions
 apparently
been cosets only).  We shall suggest that while the master equation has only
`irreducible'  solutions
(like  standard  WZW and cosets)
 our equation   contains also `reducible' ones (like chiral gauged WZW  and
its generalisation equivalent to $[G\times H_c]/H_c$ coset model). It should be
possible to
represent the  latter  (using chiral  combinations of original non-chiral
currents) as direct
products  of `irreducible' solutions.
 Solutions of  our equation   correspond  to field theories that are manifestly
Lorentz
invariant off the conformal point.  At the same time,  the  action
that corresponds to  the  off conformal point extension of the large $k$ limit
of the Hamiltonian of
a {\it  generic}  affine-Virasoro construction  is not Lorentz invariant
\hally\
(see  also \halpe). We shall suggest    a class  of non-Lorentz-invariant \sms
that may  have their conformal invariance conditions  being  related  to the
master
equation.

 %%%%%%%%%%%%%%%%%%%%%%%%%%%%
\newsec {The model}
%%%%%%%%%%%%%%%%%%%%%%%%%%%
The model we shall study below  can be represented as   the following
generalisation of
the gauged WZW model $(S_Q=k I_Q)$
$$ I_Q (g,B) = I(g)  +{1\over \pi }
 \int d^2 z \Tr \bigl[ - B\bd g g\inv   +
  g\inv\del g \B + g\inv Bg \B +   B(Q-2I)\B    \bigr] $$
$$ = I(g)  +{1\over \pi }
 \int d^2 z \Tr \bigl[ - B^a \J_a  +
 \B^a J_a + B^a ( C_{ab} +  Q_{ab} - 2 \eta_{ab} ) \B^b   \bigr] \ , \eq{2.1}
$$
where $a,b,...$ are the indices from the Lie algebra of a group $G$ (which we
shall assume
to be simple)
and\foot{We shall use the following conventions: $ [T_a,T_b]= if^c_{\ ab} T_c\
, \ \
f_{a bc}f_{abd} =  c_G \eta_{cd}\ , \ \  f_{abc}= \eta_{ad} f^d_{\ bc} $.
$\eta_{ab} = \d_{ab}$ in the
case of a compact group.
 We shall use $\eta_{ab}$ to raise and
lower indices and to contract repeated indices.  Depending on a context, both
$\eta_{ab}$  and
$\d^a_b$  will be denoted by   $I$.  We shall assume that $(ab) = \ha (ab +
ba)\ , [ab] = \ha (ab -
ba)$ , etc.  We shall use the same letters $G$ and $H$ for the algebras of $G$
and $H$. $H_c$ will
denote the maximal abelian (Cartan) subalgebra of $G$, $\ \dim H_c = r$.}
 $$
 I \equiv  {1\over 2\pi }
\int d^2 z  \Tr (\del g^{-1}
\bd g )  +  {i\over  12 \pi   } \int d^3 z \Tr ( g^{-1} dg)^3   \  , \
\eq{2.2} $$
$$
  g\inv\del g = J^aT_a \ , \ \ \   \bd g g\inv= \J^aT_a  \ , \ \ \
J_a = \eta_{ab} J^b \ , \ \ \J_a = \eta_{ab} \J^b\ , \ \ B= B^a T_a\ , \ \ \B=
\B^a T_a\ ,
$$ $$
\Tr (T_a T_b) = \eta_{ab}\ , \ \ \  C_{ab} \equiv  \Tr (T_a gT_b g\inv)\ , \ \
\ C^TC=I \ , \ \
(C)^a_{\ b} = \eta^{ad} C_{db}  \ .
\eq{2.3} $$
 The model is parametrised by  a constant   matrix $Q_{ab}= \Tr(T_a Q T_b)$.
In what follows we shall take $Q $ to be symmetric though it would be
interesting to generalise
the discussion by relaxing this assumption.

We shall also assume that the vector field $(B,\B)$ has the form (this will be
important in the case
when $Q_{ab}$ is
 degenerate)
$$ B^a = Q^a_{\ b} A^b \ , \ \ \ \B_a =  Q^a_{\ b} \A^b \ ,  \ \ \  Q^a_{\ b}=
\eta^{ac} Q_{cb}\ ,
\ \ \ Q^T=
Q \ .  \eq{2.4} $$ Then the $B(...)\B$-term in (2.1) takes the form
$$  A_a M^a_{\ b} \A^b \ , \ \ \ \  \  M \equiv  Q(C +  Q  - 2I) Q   \ .
\eq{2.5} $$
The standard WZW theory corresponds to the limit $Q\ra \infty I$ when $B,\B$
decouple.
The  action of the  gauged $G/H$ WZW model (with a vector subgroup $H$) is
recovered  when
$$ { \rm  \ gauged \  WZW \ (vector) \ : \ \ \ }  Q = P \ , \ \ \ \ P^2 = P \ ,
\eq{2.6} $$
where $P_{ab}$ is the projector on the Lie algebra of the subgroup $H$.
The case when $H$ is the axial subgroup (which is not anomalous if $H$ is
abelian, see e.g. \gin)
corresponds to  $$ { \rm  \ gauged \  WZW \ (axial) \ : \ \ \ }
Q = 3P   \ . \eq{2.6'}  $$
Another special limit of (2.1) is the  chiral  gauged $G/H$  WZW model
\tye\gin\sfet\ST\
$$ {\rm chiral \ gauged \  WZW \ :\ \ \  } Q = 2 P   \ ,  \eq{2.7} $$
 when  the matrix $M$ in (2.5) is equal simply to $PCP$.  The model (2.1) with
$\Qab = (a +1) P_{ab}$ where $a$ is a number was discussed in \ST\STT.

Both the gauged and chiral gauged WZW models are conformal invariant
at the quantum level since their actions can be represented as
combinations of independent WZW actions
$$ I(g,A) = I (h\inv g \bh ) -  I (h\inv \bh) \ , \ \ \
I_{chir} (g,A) = I (h\inv g \bh ) -  I (h\inv ) - I(\bh) \ ,
 \eq{2.8} $$
 where
 $A$ and $\A$  have been  parametrised in terms of $h$ and $\bh$  which
take values in  $H$, $\  A = h \del h\inv   , \ \  \A = \bh
\bd \bh\inv  . $
In what follows we shall find  (in the one-loop approximation)
the conditions on the matrix
$Q$ under which  the action  (2.1) describes a conformal  theory.

Solving for $A,\A$ in (2.1),(2.4) one finds  the following semiclassical action
for the group
variable $g$
$$ I_Q (g) = I(g)  + {1\over  \pi }
 \int d^2 z \   \M_{ ab}    J^a \J^b \ , \eq{2.9} $$
$$
\M \equiv  Q M^{-1} Q = Q [Q(C +  Q  - 2I) Q ]\inv Q
\ ,  \eq{2.10} $$
where the inverse is defined  on the subspace on which $Q$ is non-degenerate.
Note that the matrix $\M$ is  a non-trivial function of $g$ and (2.9) (as well
as (2.1))
does not have  a global $G$-invariance.
The action (2.9) can be  interpreted as a WZW action  perturbed by an operator
which is
not integrably marginal \chaud\ in general. It can be put in the form of an
integrably marginal
${\cal J }{\bar {\cal J}}$-operator  in the case when
$Q$  will have  the form $P\r P$ where $P$ is the  projector on the Cartan
subalgebra of $G$
(in agreement with the previous discussions \hussen\nap\kiri\givkir).
Such $Q$ will be one of the solutions of the conformal invariance conditions of
the model
(2.9)
to be derived below.

It should be stressed that our true starting point is the action (2.1) while
(2.9)
appears only  as a semiclassical  approximation. For example, to preserve
conformal invariance
present in (2.1) for special $Q$  (2.9)
should be supplemented by the dilaton coupling originating from the integral
over $A$ in (2.1).
  When $Q $ is non-degenerate the
action (2.9) is  related   to the  action discussed in \kart\  in connection
with bosonisation
\witt\boul\ of the non-abelian Thirring model \dash.\foot{As was shown in
\kart\ the bosonised
version of the Thirring model with the coupling matrix $S$ can be represented
(in the sense of
equivalence of the corresponding generating functionals for currents)
as the  `deformed' WZW model (2.9)
with a non-polynomial  dependence of $\M$ on $S$. The relation of  (in general,
non-symmetric) $Q$
 to $S$ in \kart\ is the following: in  the `left-right decoupled' scheme $Q-2=
\ha [{S/
2\pi}-   ({S/2\pi})\inv]$ (the transformation to other schemes is implemented
by  shifting $Q$ by a
constant times a unit matrix, e.g. $Q\ra Q +I$ in the vector scheme).  Note
that the duality
symmetry  in \kart\  ${S/ 2\pi}\ra -   ({S/2\pi})\inv$  is trivial  in terms of
 $Q$. As we shall
note  below (see (2.19),(2.20)), there is   also a non-trivial duality-type
symmetry  ${S/ 2\pi}\ra
({S/2\pi})\inv$  which relates two `dual' models of the type (2.9) (see also
\kut).  In the next
section  we  shall derive the condition of conformal invariance of this model
(to the leading order
in $1/k$ but to all orders in the (symmetric) coupling)  as well as an `action'
(`potential')  from
which it follows,  thus  providing a natural extension of the discussion  in
\kart. It should be
emphasized that our approach is more general than that of  \kart\  being
applicable  also when
 $Q$   is  degenerate so we are able   to include systematically  the cases of
the $G/H$ gauged (and
chiral gauged) WZW models. In these cases it is important also to take into
account the dilaton
coupling (see (3.1),(3.4)) that in general should be added to (2.9). }  In this
case  $\M$
takes the form   $$ \M =  ( C + Q - 2 \I)\inv \    \ .   $$
The WZW action corresponds to $Q=\infty I$.    When
$Q=2\I$   (i.e. in the case of the $G/G$ chiral gauged WZW model, cf. (2.7))
 eq.(2.9) takes the form of the WZW action with the sign of the first term in
(2.2)
reversed.\foot{In the context of ref. \kart\  this case corresponds to the
second (`dual') conformal
point \dash\ of the Thirring model (see also \kut). }
\def \dx {{\dot x}}

It is straightforward to determine  the classical Hamiltonian (or
`00'-component of the stress tensor)
that corresponds to  (2.9).  Starting with (2.1),(2.4)  we  define  the momenta
$p_m$ as derivatives
of the Lagrangian over ${\dot x}^m$ where $x^m$ are the group space
coordinates.
 Using Minkowski notation we have
for the pure WZW action
$$I(g) =   {1 \over  4 \pi  } \int d^2 z  \bigl[ \ha { G}_{0 mn }  (x)
(\dx^m\dx^n  - x'^m x'^n)  +
 B_{0 mn} (x) \dx^m x'^n \bigr]
 \ , \eq{2.11} $$ $$    g\inv  dg = iT_a E^a_m (x)  d x^m \ , \ \ \ \  \  dg
g\inv = iT_a \E^a_m (x)
dx^m \ , $$ so that $p_m = {1 \over  4 \pi  } ( { G}_{0 mn }  \dx^n  +  B_{0
mn}  x'^n) $.
Assuming that $p_m$ and $x^m$ have standard Poisson bracket relation one finds
that
the currents $$J_a= J{ - a }(x,p) =  E_{am} (x)  (\dx^m - x'^m) = 4\pi E^{m}_a
(p_m - {1\ov 4 \pi }
B_{0mn} x'^n)  - E_{am} x'^m  \ ,  $$
 $$  \J_a= J_{ + a} (x,p)   = T_a \E^a_m (x)  (\dx^m  + x'^m) = 4\pi \E^{m}_a
(p_m - {1\ov 4
\pi } B_{0mn} x'^n)  +  \E_{am}x'^m  \ ,  $$
form two commuting affine algebras  \witt\bow.  In the case of (2.1)  the
expressions for
momenta   contain extra  terms linear in $B,\B$ or $A,\A$.  The currents that
form the affine
algebras are again given by the same functions of $x^m$ and new $p_m$  as in
the WZW theory.
Using the same notation $J, \ \J$ for these currents one finds for the
Hamiltonian as function of
independent phase space variables $x,\ p, \ A, \A$
  (cf. \bow\ST)
$$\H=  \ha    J^2 + \ha {\bar J}^2  - 2 J \B + 2 \J  B    + 2 B ( Q -2\I)  \B
+  B^2 +   \B^2 \ .
\eq{2.12} $$
Elimination of  $A,\A$  defined in (2.4) gives (note that $J^2=\J^2$)
$$ \H  = \ha L^{ab} (J_a  J_b  + \J_a \J_b) +  \L^{ab} J_a\J_b  \ , \eq{2.13}
$$
$$ L = \I +  2 QF\inv Q \ , \ \ \  \L= 2  Q(Q-2\I) F\inv Q \ , \ \ \  \ \
F\equiv Q(Q-3\I)(Q-\I)Q \ . \eq{2.14} $$
In the  singular  case of the gauged WZW model  (2.6) one is to take  the limit
$Q= P + \epsilon, \ \epsilon\ra 0 $, so that the resulting  Hamiltonian is
finite on
the  subspace where $J_H-\J_H=0$  ($J_H=PJ, \ \J_H = P\J$)
\bow\ST\foot{Similar result is found in
another singular limit  (2.6$'$) $Q= 3P + \epsilon, \ \epsilon\ra 0 $:
$ \H = J^2  +  J_H\J_H
-  {1\ov 2 \epsilon } (J_H +\J_H)^2 $, i.e. $ \H \ra   \H = \ha(J^2 - J_H^2) +
\ha(\J^2 -
 \J_H^2) \ $  on a subspace where $J_H +\J_H=0$. }
 $$ \H = J^2  -  J_H\J_H
-  {1\ov 2 \epsilon } (J_H-\J_H)^2\  \ra  \ \H = \ha(J^2 - J_H^2) + \ha(\J^2 -
 \J_H^2) \  . \eq{2.15}
$$
In the chiral gauged WZW model case (2.7)  \sfet\ST
$$ \H_{chir} =  \ha (J^2-2J_H^2) + \ha (\J^2-2 \J_H^2)  \ . \eq{2.16} $$
Note that the  `non-diagonal' $J\J$ term in (2.13) is {\it non-vanishing} in
all other cases when
$Q^2\not=2Q$.

Introducing the matrix $K$
 $$ K\equiv{ Q^3-3Q^2\ov Q^3-Q^2} \ , \eq{2.17 } $$
we get
$$L = \ha (K +K\inv) \ , \ \ \ \ \L= - \ha (K -K\inv) \ , \eq{2.18} $$
so that (2.13) can be put into the  form
$$ \H= \fourth K^{ab} (J_a -\J_a)(J_b -\J_b) + \fourth K^{-1 ab} (J_a
+\J_a)(J_b +\J_b)
\ ,  \eq{2.19} $$
which has  an obvious duality-type symmetry
   $$K\ra K\inv\ , \ \ \ \ \  \J\ra -\J \  . \  \eq{2.20}
$$
Note that (2.20) is not a symmetry of the action (2.9) itself,  but it should
 relate   two  `dual' actions which give  the same generating functionals for
the correlators of the corresponding currents (as in the simple case of a
single  coupling in \kut).\foot{The transformation   $K\ra - K\inv , \ \  \J\ra
-\J $ is also a
symmetry if it is accompanied by  reversing  the sign of the coefficient $k$ of
the action (for
a discussion of a similar symmetry in the quantum generating functional of the
single-coupling
Thirring model  see \kut).  In the case of a non-degenerate $Q\ $ the matrix
$K$ is related to the
Thirring coupling of \kart\kut\ by  $Q-2 = - {K+1\ov K-1} = \ha [{S/ 2\pi}-
({S/2\pi})\inv] $ and so
$K\ra K\inv$ corresponds to ${S/ 2\pi}\ra   ({S/2\pi})\inv$. } In particular,
this  should be   the
case for $K=1+ P\r P$  (with $P$ being the projector on the Cartan subalgebra
of the algebra of $G$)
when (2.20) should be related to a particular element of the  $O(2r,2r)$  ($r=
{\rm rank \ } G$)
duality group (see \givroc\hussen\nap\kiri\givkir).

When $Q$ is non-degenerate we can represent the action (2.1) in the form
$$
I_Q(g, h, \bh) =  I ( \tilde g ) + I'(h, \bh) \ , \ \ \eq{2.21}    $$
$$I'=  -  I (h\inv ) - I(\bh)
 +{1\over \pi }
 \int d^2 z \Tr \bigl[  h \del   h\inv  (Q-2) \bh\bd \bh\inv  \bigr]\ ,
 \eq{2.22} $$
 where
$$ \tilde g = h\inv g \bh\ ,  \ \ \
 B= h \del h\inv \  , \ \  \B = \bh\bd \bh\inv  \ .  \ \  $$
By formal manipulations in the path integral and use of the Polyakov-Wiegmann
 formula  the action $I'$
can be  transformed into (we shall ignore the quantum shifts of  $k$, see \kut)
$$ I'(h, \bh) \ra     -  I (h\inv ) - I(\bh)  +   {1\over \pi }
 \int d^2 z \Tr \bigl[ - B'\bh \bd \bh\inv   +
  h\del h\inv  \B' +    B'(Q-2)\inv \B'    \bigr] \ra  $$
$$ I''=  -  I ({h'}\inv ) - I(\bh') + I(u\inv) + I(\bar u) + {1\over \pi }
 \int d^2 z \Tr \bigl[  u \del   {u}\inv  (Q-2)\inv  {\bar u}\bd {\bar u}\inv
\bigr]\ ,
 \eq{2.23} $$
$$
B'= u \del u\inv \  , \ \  \ \B' = \bar u  \bd {\bar u}\inv \ , \ \ \
{h'}= {\bar u}\inv h \ , \ \ \ \ \bh'= u\inv  \bh  \ . $$
Integrating out $h', {\bar h}'$  and redefining $g$  we get the  dual action
$\tilde I_{Q'} (g', u,
\bar u) $ with $k\ra -k$ and
$$ Q'-2= - (Q-2)\inv\ , \ \ \ \  K'= {Q'-3\ov Q'-1} = - K\inv \ .  \eq{2.24} $$

%%%%%%%%%%%%%%%%%%%%%%%%%%%%%%%%%%%%%%%%%%%%%%%%

%%%%%%%%%%%%%%%%%%%%%%%%%%%%%%%%%%%%%%%%%%%%%%%%%%%%%%%%

\newsec{Sigma model representation and equations  of conformal invariance}

%%%%%%%%%%%%%%%%%%%%%%%%%%%%%%%%%%%%%%%%%%%%%%%%%%%%%%%%%%%
As in the cases of gauged and chiral gauged WZW models (see e.g.
\bs\aat\sfet\ST)
one can represent
the semiclassical action (2.9) or $S_Q(g)= k I_Q(g)$ in the \sm form
$$S_Q(x) =   {1 \over \pi \a' } \int d^2 z  ({ G}_{mn } + B_{mn}) \del
x^m\bd x^n   +   {1 \over 4\pi  } \int d^2 z \sqrt \g R^{(2)} \p \ , \ \  \ \a'
= {2\ov k } \ ,
\eq{3.1} $$
 $$ G_{mn}  = { G}_{0 mn }  - 2\M_{ac} E^a_{(m }\E^c_{n)}= g_{ac}(x) E^a_m
E^c_n\ , \eq{3.2} $$
$$  B_{mn}=  B_{0mn} -   2\M_{ac} E^a_{[m }\E^c_{n] }= b_{ac}(x) E^a_m E^c_n \
, \eq{3.3} $$
$$ \p = \p_0 - \ha {\rm ln}\det M  \ . \eq{3.4} $$
We have introduced the  coordinates on the group $G$ and the vielbein $E^a_M$
according to\foot{We
rescale the coordinates $x^m$ to make them dimensionless,
absorbing the `radius' of the group space into $\a'$. }  $$
g\inv \del g = iT_a E^a_m (x)  \del x^m \ ,  \ \ \  \del g  g\inv  = iT_a
\E^a_m (x)  \del x^m \ , \
\ \  \E^a_m =  C^a_{ \ b } E^b_m \ . \eq{3.5}  $$
$G_{0mn}$ and $B_{0mn}$ are the WZW couplings corresponding to the group space
$$ G_{0mn} = \eta_{ac} E^a_m E^c_n \ , \   \ B_{0mn} = b_{0ac} E^a_m E^c_n \ ,
\ \
 \ H_{0mnk}= 3\del_{[k} B_{0mn]} = - f_{abc}E^a_mE^b_nE^c_k
  . \eq{3.6}  $$
We have also included the dilaton coupling that originates from the determinant
of integration over
$A,\A$ in passing from (2.1),(2.5)  to (2.9).  As in the gauged WZW case the
presence of a
non-trivial dilaton is related to the fact that $\det G\not= \det G_0$, i.e.
the dilaton can be
also represented in the form
$$ \p =\p_0  + \fourth {\rm ln}\det {G\ov G_0} = \p_0  + \fourth {\rm ln}\det g
\ . \eq{3.4'} $$
 The  matrix functions $g_{ac}$ and $b_{ac}$ in (3.2), (3.3)  have the
form (see (2.5),(2.10)) $$g = I - QM\inv Q C  - C^TQ M^{-T}Q \ , \ \ \ b=b_0 -
QM\inv Q C  + C^TQ
M^{-T}Q \ . \eq{3.7} $$ In the gauged WZW case (2.6) when $Q^2=Q=P$ the metric
(3.2) is degenerate
having $\dim H$ null vectors,  $G_{mn} (\E^m_a -E^m_a)=0$ or $g
(C^T-1)P=0$.\foot{Similar degeneracy
takes place in the axial gauging case when  $Q=3P$,
 $\ g(C^T+1)P=0$.}

In what follows we shall determine the conditions on $Q$ under which the model
(3.1) is conformal
invariant.
 The one-loop equations   of conformal invariance of the \sm  (3.1)  have the
standard form \call
 $$   {\bar \b}^G_{mn} = R_{mn} - \fourth H_{mkl}H_n^{\ kl} + 2D_m
D_n \p  = 0 \ , \eq{3.8} $$ $$ {\bar \b}^B_{mn}  = - \ha D_l H^l_{\ mn}  +
H^l_{\
mn}\del_l \p =0 \ ,  \eq{3.9} $$
$$ {\bar \b}^\p =   {1\ov 6} (D-\C)   + \a' [ - \ha  D^2 \p +
(\del \p)^2 -  {1\ov 24} { H}^{2}_{lmn  }] =0 \ ,   \eq{3.10} $$
where $D= \dim G$  and $\C$ is the total central charge. Using (3.8),(3.10),
i.e.
$$   {\tilde \b}^\p={\bar \b}^\p - \fourth  {\bar \b}^G = {1\ov 6}(D- \C ) +
O(\a') =0 \ ,
\eq{3.10'}  $$
  $\C$   can be represented in the form
 $$ \C  = D  - {3\ov 2 }
\a' [ R - {1\ov 12} { H}^{2}_{lmn  }  + 4 D^2\p -4(\del \p)^2] + O(\a'^2) \ .
\eq{3.11} $$
 $G_{mn}$ and $B_{mn}$ depend on $x$ only through $E^a_m$ and $C_{ab}$ so it
 is straightforward to
compute the curvature of $G_{mn}$ and $H_{mnk}$ with the help of  the relations
$$ \del_m E^a_n - \del_n E^a_m = - f^a_{\ bc} E^b_m E^c_n \ , \ \ \  \ \
\del_m C_{ab} = - C_{ac}f^c_{\ bd} E^d_m   \ , \eq{3.12} $$
i.e. $R^m_{\ nkl}$ and $H_{mnk}$  will be given by  sums of products of
$E^a_m$, $f^a_{\ bc}$ and
matrix functions of $C_{ab}$ and $Q$.
That means that the geometrical objects   appearing  in (3.8)--(3.10)  are
`dimensionally reducible' (cf. \coq) in
the sense that their expressions at an arbitrary point of the group space $G$
can be determined (by integrating differential equations  implied by (3.12))
 from their
values  at  a  particular point of $G$.

To establish  the conditions on $Q$ that follow from (3.8),(3.9) one can,
therefore,  use a short-cut
method by expanding  $ E^a_m$ and $C_{ab}$ in normal coordinates near the
unit element of the group (see e.g. \schouten)
$$ E^a_m = \big({\e{f} - 1 \ov f } \big)^a_m = [I + \ha f + {1\ov 6} f^2 +
O(x^3) ]^a_m \ , \eq{3.13}
$$
$$ { g= \exp ({iT_a x^a }) }\  , \ \ \
(f)^a_{\ b }\equiv f^a_{\ bc }x^c \ , \ \ \  f^T=-f\ , \ \ \ E^a_m
x^m=x^a\ , \eq{3.14} $$ $$
 C^a_{\ b} = ({\rm e}^{-f})^a_{\ b} = [I - f + \ha f^2 + O(x^3) ]^a_{\ b}  \  .
\eq{3.15} $$
Then the matrix $M$ in (3.6),(2.5) is  given by
$$ M = Q^3-Q^2  - QfQ + \ha Qf^2Q +  O(x^3) \ , $$
and so
$$g=K + \ha(fK-Kf) -\fourth fKf + \fourth KfKfK + O(x^3) \ , \eq{3.16}  $$
where the constant matrix $K_{ab}$ have already appeared in (2.17), i.e.
$$ K\equiv I-2 Q (Q^3-Q^2)\inv Q = (Q^3-3Q^2)(Q^3-Q^2)\inv \ . \eq{3.17 } $$
The inverse is assumed to be defined on a subspace where $Q^3-Q^2$ is
non-degenerate.
When $Q$ is non-degenerate
$$  K= {Q-3I\ov Q-I} \ , \ \ \ \ \ \ \ Q= {K-3I\ov K-I} \ . \eq{3.17'} $$
   The expansions of $G_{mn}$,  $B_{mn}$ and $\p$ have the form
$$ G= K - {1\ov 12} (Kf^2 + f^2 K)  + \fourth KfKfK + O(x^3) \ , \eq{3.18  } $$
$$ B=B_0 + \ha f - \ha KfK + O(x^2)\ , \eq{3.19 } $$
$$\p = \p_0  - {1\ov 16}\Tr ( f^2 - KfKf) +  O(x^3) \ ,  \eq{3.20  } $$
so that  one finds for the Ricci tensor, $H_{mnk}$ and $D_mD_n \p$
$$R_{mn} = \fourth [ - f_{mkl}f_{nkl}  + f_{mkl}f_{n\unk\unl}
+ f_{\unm kl}f_{\unn \unk \bl}
+ f_{kl(m}f_{\unn ) k\unl}  -  f_{kl(m}f_{\unn ) k\bl}] + O(x)\ , \eq{3.21} $$
$$ H_{mnk} = \ha ( f_{mnk} - f_{m\unn  \unk} - f_{\unm n\unk} - f_{\unm \unn
k}) + O(x) \ , \eq{3.22}
$$ $$ D_m D_n \p = {1\ov 8} ( f_{mkl}f_{nkl} -  f_{mkl}f_{n\unk\unl}) + O(x) \
, \eq{3.23} $$
where the repeated  indices are contracted with $\eta_{ab}$,
underlined index  indicates an extra factor of $K$ and an index with a bar -- a
factor
of $K\inv$, for example,
$$ f_{{\underline a}bc} = K_{aa'} f_{a'bc}\ , \ \ \ \ \ f_{{\bar a}bc} =
K^{-1}_{aa'}
f_{a'bc}\ . \eq{3.24} $$
We have used that $f_{abc}=\eta_{ad}f^{d}_{\ bc}$ is totally antisymmetric.

The above expressions reduce to the well-known results in the group space limit
when
$Q=\infty I$, i.e. $K=I$,
$$R_{mn} = \fourth  f_{mkl}f_{nkl}\ , \ \ \  H_{mnk} = - f_{mnk}\ , \ \ \p=\p_0
\ . \eq{3.25} $$
In the case of the gauged WZW model (2.6) one should first project the metric
$G_{mn}$
and $H_{mnk}$  on $G/H$ to make them non-degenerate and only then expand in
powers of $x$.
Then eqs. (3.13)--(3.18)  still apply if all uncontracted indices are
multiplied by
the $G/H$ projector $P^{\perp}=I-P$ and  both  $K$ and $K\inv$ are   replaced
by  $P^{\perp}$
$$ K= I- P = P^{\perp} \ , \ \ \  K^{-1}=P^{\perp}\  , \ \  \ \ P^{\perp 2
}=P^{\perp}\ .  \eq{3.26}
$$ In the case of the chiral gauged WZW model (2.7) one finds that $K=K\inv$,
i.e.
$$ K = I-2P  \ , \ \ \ \ K\inv = I-2P  \ , \ \ \ \ K^2=I \ ,  \eq{3.27} $$
so that it is not necessary to distinguish between the $\unm$ and $\bm$
indices.
If $G$ and $H$ are  compact, $K$ in (3.27)  can be put into the form
 $K_{ab}= {\rm diag} (+1, ..., +1, -1,..., -1)$.\foot{ Note that $K_{ab}$ is
the constant part
of the target space metric (3.18) (cf. also the Hamiltonian (2.19))
 and thus the signature of it is
determined by the signatures of the Killing metrics of $G$ and $H$. For
example, we get
just one time-like  direction if $G$ is compact and $H=U(1)$ (see also \ST). }

Since the objects in  eqs.(3.8)--(3.10) are  expressed in terms
of  products  of the  universal  quantities  $E^a_m,\  f_{abc} $ and algebraic
functions of $ C_{ab}$
they should  be  satisfied automatically if satisfied at  $x=0$.
 Computing the constant part of
eq.(3.8) using (3.16)--(3.18) we get  the following basic equation on the
matrix $K$  (or  on $Q$)
$$   ({\bar \b}^G_{mn})_{x=0} = {1\ov 8} (  f_{mkl}f_{nkl}  -\ha
f_{mkl}f_{n\unk\unl}
-\ha  f_{mkl}f_{n\bk\bl}
+ f_{\unm kl}f_{\unn\unk\bl}  -  f_{\unm kl}f_{\unn kl}) =0 \ . \eq{3.28} $$
The  trace of this equation  with $K\inv_{mn}$  is
$$ f_{mkl}(f_{\bm kl} - f_{\unm kl} +\ha  f_{\bm\unk\unl}
-\ha  f_{\bm\bk\bl} ) =0 \ . \eq{3.28'} $$
The dilaton equation (3.10) implies
$$ \C = D - {1\ov 16}\a' (f_{mkl}f_{\bm\bk\bl} -3 f_{mkl}f_{\unm\unk\bl}
+6f_{mkl}f_{\unm kl}) + O(\a'^2) \ , \eq{3.29} $$
while combining (3.28$'$) and (3.29) one gets the expression that follows from
(3.11)
$$ \C = D - {1\ov 32}\a' (-f_{mkl}f_{\bm\bk\bl} -3 f_{mkl}f_{\unm\unk\bl}
+6f_{mkl}f_{\unm kl} + 6f_{mkl}f_{\bm kl}  ) + O(\a'^2) \ . \eq{3.30} $$
This representation is of interest since, in agreement with  general
expectations,
one can check that $\C (K)$ in (3.30) plays the role of  an `action' for the
equation
(3.28). As is well known, eqs.(3.8)--(3.10) follow from the effective action
$S= \int d^Dx{\sqrt G}
{\rm e}^{-2\p} {\tilde \b}^\p $, or, up to a shift in the constant term in
(3.10), from
$$S= \int d^Dx\ {\sqrt G}\ {\rm e}^{-2\p}\  \C  \ , $$
 with $\C$ given by (3.11). For the background under
consideration  the `measure factor' $ {\sqrt G}{\rm e}^{-2\p} =  {\sqrt G_0}$
is  $K$-independent
and that is why   $ \del \C / \del K = 0$ is equivalent to
(3.28).\foot{Eq.(3.30) gives the
`effective potential' (of the structure $K\inv K\inv K\inv ff + KKK\inv ff+ Kff
+ K\inv ff$)
 for the coupling $K_{ab}$. A similar  cubic  $SSSff$ potential  for the
coupling $S$ of the Thirring
model appeared  in   \thirpot.}

Eq.(3.9)  is always satisfied to  the leading order. The constant term in (3.9)
should be given by a sum of the products of two factors of $f_{abc}$, one
$K\inv$ and  several
$K$'s but  an  antisymmetric tensor of such  structure  does not exist if
$f_{abc}$ is totally
antisymmetric and $K_{ab}$ is symmetric. The absence of an extra antisymmetric
constraint coming
from (3.9) is consistent with the fact that our background is parametrised by a
symmetric matrix
$K_{ab}$ (which is the variable in  the `action' (3.30)).

Let us emphasize the  central  role played by the matrix $K_{ab}$ (3.17) in the
above  construction.
This is clear from the representation (2.19) for the Hamiltonian  (2.13),(2.18)
which  is of course  related  to the following expressions for   the inverse
matrices to
$g_{ab}$ and  $G_{mn}$ in (3.2),(3.7)
$$ g\inv = \ha ( L + C^TLC + \L C  + C^T\L ) = \fourth [ (C^T-1)K(C-1) +
(C^T+1)K\inv (C+1) ]
\ , \eq{3.31}
$$
$$ G^{mn} = \ha L^{ab}(E^m_aE^m_b + \E^m_a\E^m_b)  + \L^{ab} E^m_a\E^m_b $$ $$
 = \fourth K^{ab} (E^m_a -\E^m_a)(E^m_b -\E^m_b) + \fourth K^{-1 ab} (E^m_a
+\E^m_a)(E^m_b +\E^m_b)
\ , \eq{3.32} $$
which are invariant under (2.20), or
$K\ra K\inv\ ,  \  \ \E\ra -\E. $

%%%%%%%%%%%%%%%%%%%%%%%%%%%%%%%%%%%%%%%%%%%%%%%%
\newsec{Solutions of the  conformal invariance condition }
%%%%%%%%%%%%%%%%%%%%%%%%%%%%%%%%%%%%%%%%%%%%%%%%%%
Let us now study possible solutions of the conformal invariance equation (3.28)
or, explicitly, of
$$      f_{mkl}f_{nkl}  -\ha  f_{mkl}f_{nk'l'} K_{kk'}K_{ll'}
-\ha  f_{mkl}f_{nk'l'} K^{-1}_{kk'}K^{-1}_{ll'} $$ $$
+  K_{mm'}K_{nn'} f_{m' kl}f_{n'k'l'} K_{kk'}K^{-1}_{ll'}
 -   K_{mm'}K_{nn'} f_{m' kl}f_{n' kl} =0 \ .
\eq{4.1} $$
In general,  this equation is
not  invariant under $K\ra K\inv$. However, the `coupling constant duality'
$K\ra K\inv$  becomes
manifest in the $\b$-function in the simplest possible case  when $K$ is
proportional to  a unit
matrix$$
K = K_0I \ , \ \  \ \ \ \ \ Q= {K-3I\ov K-I}= {K_0-3\ov K_0-1} I \ . \eq{4.2}
$$
 Then (3.28) and (3.30) take the form\foot{Similar  expressions (which are
exact in $K_0$ but
first order in $1/k$)  for the $\b$-function
and the  central charge `potential' were derived
in the Thirring model context
 \dash\thir\ in \kut. In our approach  (4.3)  follows simply  from the known
\sm $\bar\b$-function
(3.8).  Our result for the  `potential' (4.4) or  (3.11) is not duality
symmetric but,
in  general,   the `potential' is  not defined unambiguously out of the
conformal point.
It should be noted also that in (4.3),(4.4) we have included the dilaton
contribution
(the expression
for the dilaton (3.4),(3.20) is non-trivial for $K^2_0\not=1$ even in the
one-coupling case)
 which was not considered in \kut\  where the Thirring model was the starting
point.}
 $$
({\bar \b}^G_{mn})_{x=0} = -{1\ov 16} ( K_0- K_0^{-1})^2 c_G \eta_{mn} +
O({1\ov k} )  \ , \eq{4.3}
$$
 $$ \C = D - {1\ov 16 k } (-K_0^{-3} + 3K_0+ 6K_0^{-1} )c_G D  + O({1\ov k^2} )
\ . \eq{4.4} $$
The only two conformal points are $K_0=1$ or $Q=\infty I$ (WZW model) and
$K_0=-1$ or
$Q=2I$ (WZW model with the  reversed  sign of the first term,
 or  $G/G$ chiral gauged WZW model).  For $K_0=1$  the central charge (4.4)
takes the standard form
$$ \C  = D -   {1\ov 2k} c_G D +  O({1\ov k^2} )
= {Dk\ov k+ \ha c_G} \ . \eq{4.5} $$
A less trivial  solution corresponds to the   $G/H$ gauged WZW model
(2.6),(3.26)  when
$K=P^{\perp}=P_{G/H}$.
In general,
 if
$K=K\inv$, eq.(4.1) reduces to
$$    ( \eta_{mm'}\eta_{nn'}
-  K_{mm'} K_{nn'}) ( \eta_{kk'} \eta_{ll'}-   K_{kk'}K_{ll'})  f_{m' kl} f_{n'
k'l'}
=0 \ , \ \ \  K=K\inv \ .   \eq{4.6} $$
Furthermore,  if we formally  take  $K$ to be  a projector, i.e. $K=K\inv=K^2$
then using again the
notation in which the indices    multiplied  by $K$ are underlined,  we get
from (4.6)
(cf.(3.28)) $$     f_{mkl}f_{nkl}  -   f_{m\unk\unl}f_{n\unk\unl}
+ f_{\unm \unk\unl}f_{\unn \unk\unl}  -  f_{\unm kl}f_{\unn kl} =0 \ , \ \  \ \
 K=K^2 \ .  \eq{4.7}
$$ The $KK$-projection (i.e. the product with $K_{mm'}K_{nn'}$) of this
equation is satisfied
automatically while the projections  $KK^{\perp}$  and  $ K^{\perp}K^{\perp}\ $
 $(K^{\perp}=I-K) $
give the following two equations
 $$       f_{{\hat m }\unk\unl}f_{\unn\unk\unl}= f_{{\hat m }kl}f_{\unn kl}
= c_G \eta_{ {\hat m }\unn } =0 \ ,  \eq{4.7'} $$ $$
\  f_{{\hat m }\unk\unl}f_{{\hat    n}\unk\unl}= f_{{\hat m }kl}f_{{\hat
n}kl}=
       c_G \eta_{{\hat m }{\hat    n}} \ ,  \eq{4.7''}  $$
where  $\hat m $ denotes an index projected with the help of $K^\perp$.

In the case of the gauged WZW model  the gauge invariance implies that one
should consider only
the   $KK$-projection  of (4.1), i.e. (4.7$'$),(4.7$''$) are absent and thus
(4.1)
is satisfied.\foot{ Note that  our derivation of (4.1) from (3.8)  formally
applies only when the
metric (3.18) or $K$ is non-degenerate. The gauged WZW model case is special,
having explicit gauge
invariance that implies the use  of  appropriate projectors (or gauge fixing).}
Then  the expression for the central charge
(3.29) becomes\foot{Because of gauge invariance of the path integral of the
gauged WZW model
\gwz\karabali\ one is also to replace the number $D$ of  the degrees of
freedom in (3.29)  by
$D_{G/H} = D_G-D_H$.}
 $$ \C = D_{G/H}  + {1\ov 8}\a' (  f_{mkl}f_{\unm\unk\unl}  -3f_{mkl}f_{\unm
kl}) + O(\a'^2)\ .
\eq{4.8} $$
Let  the  indices $r,s,t$ be  from the algebra of the subgroup $H$  (i.e. from
the
$K^\perp$-space or the same as  indices with hats) and
 the indices $\m,\n,\l,...$ be  from
from the tangent
space to $G/H$ (i.e. from the $K$-space or the same as underlined indices).
 Then one has  $f_{\m
rs}=0$  ($H$ is a subgroup)  and
  $$\ f_{mkl}f_{\unm\unk\unl}=f_{\m\n\l}f_{\m\n\l}=
 f_{mkl}f_{mkl }-
 3f_{\m\n r}f_{\m\n r} -  f_{rst} f_{rst}\  , $$ $$
\ \ f_{mkl}f_{\unm kl}=f_{\m kl}f_{\m kl}=
f_{\m\n\l}f_{\m\n\l} + 2 f_{\m \n s} f_{\m \n s}\  .  $$
As a result, (4.8) reproduces the first term in the $1/k$-expansion  of
the central charge \gko\ of the  gauged WZW (or coset) model
$$\C= D_{G/H}  - {1\ov 2k}(f_{mnk}f_{mnk} - f_{rst}f_{rst}) +O({1\ov k^2})$$ $$
=
 D_{G/H}  -    {1\ov 2k}c_G D_G
-   {1\ov 2k} c_H D_H + O({1\ov k^2}) =  {D_G k\ov k+ \ha c_G}  - {D_H k\ov k+
\ha c_H} \ ,
 \eq{4.9} $$
Next, let us  consider the solutions with $K=K\inv ,\  K^2=I.$
 This is the case  of  the chiral gauged WZW model
(2.7),(3.27).  At the `self-dual' point  $K=K\inv$  eq.(4.1) reduces to
(4.6).
It is straightforward to check
 that this equation is
satisfied if  $K=I-2P$ with $P$  being  a  projector on a {\it subalgebra}.
Namely, one should have
 $f_{\m rs} =0$ where  we  set again  $m=(\m, r)$  with  $r,s,t, ...$
corresponding
to   $P$ and  $\m,\n,\l,...$  corresponding to  $P^{\perp}\equiv I-P$.
 When $K=K\inv$ the central charge (3.29),(3.30)  takes the form similar to
(4.8)
$$ \C = D + {1\ov 8}\a' ( K_{mm'}K_{nn'}K_{ll'} f_{mnl}f_{m'n'l'}
-3 K_{mm'} f_{mkl}f_{m' kl}) + O(\a'^2) \ .  \eq{4.10} $$
For $K=I-2P$  and $f_{\m st}=0$ we get (here $D=D_G$)
$$ \C= D + {1\ov 8}\a' (-2 f_{mnk}f_{mnk}  + 4f_{rst}f_{rst} + 12f_{\m st}f_{\m
st}) +
O(\a'^2) \eq{4.11} $$
$$ =D_G -   {1\ov 2k} c_G D_G
+     {1\ov k} c_H D_H + O({1\ov k^2}) ={D_G k\ov k+ \ha c_G}  + {c_H D_H\ov k
+ \ha c_H} \ ,
 \eq{4.12} $$
Eq.(4.10) thus reproduces the $O({1/k})$ term in the expansion of the central
charge
 of the chiral gauged WZW model as given in  \ST.
Another solution   is found by reversing the sign of $K$, i.e. $K=2P-1=
1-2P^{\perp}$
(then the sign of the $O(1/k)$ term in (4.10) is  also reversed).
In the case when $K=K\inv$ the expressions (2.18),(2.19),(3.31),(3.32)
simplify,
in particular, the Hamiltonian (2.13),(2.19) takes the form
$$ \H  = \ha K^{ab} (J_a J_b  + \J_a \J_b)  \ , \ \ \ \ \ \ K=K\inv \ .
\eq{4.13} $$
Let us now  relax the condition $K=K\inv$ and look for other solutions of
(4.1). One possible ansatz
is a  generalisation of $K=I-2P$, namely,
$$ K= \g I +  P  \r  P\ , \ \ \ \ P^2=P \ , \eq{4.14} $$
where the constant $\g$ and  the constant matrix $\r_{ab}$ are to be
determined.
One  can show (e.g. by taking the $PP,\  P^{\perp}P, \
P^{\perp}P^{\perp}$-projections of (4.1))
that  the solution exists only if $(I-P)_{mm'}P_{kk'}P_{ll'} f_{m'k'l'} =0 $,
i.e. if $P$ is a
projector on a subalgebra.
%\foot{For $\g=1$ the  equation that follows from (4.1) has the
%form:
%$$f_{\unm kl} f_{n \unk l}+ f_{\unn kl} f_{m \unk l} +
%\r f_{\unm kl} f_{\unn \unk l}
%-  f_{\unm kl} f_{n \unk \unl} - f_{\unn kl} f_{m \unk \unl}
%- \r f_{\unm kl} f_{\unn \unk \unl} =0 \ , $$
%where  the  underlined indices are projected with
%the help of $P$. This equation is consistent only if $f_{m\unk\unl} = f_{\unm
%%\unk\unl}$.}
 If  the
 subalgebra  is non-abelian    then the  only solution  is
the chiral gauged WZW one (3.27), i.e. $|\g |= 1\ , \ \r = -2\g I $. If,
however,  $P$ is  a projector
on an   abelian    subalgebra   (any subalgebra of Cartan algebra $H_c$), i.e.
 $f_{mnk} P_{nn'}P_{kk'}=0,$
  then $\g^2=1$ but the matrix  $\r $ can be   {\it arbitrary}, i.e.
$$ K= I +  P \r  P \ ,  \ \ \  \ \ \  Q= {K-3I\ov K-I}=P(1-  {2 \r\inv }) P \ ,
\eq{4.14'} $$
$$ K=I + \bar \r\  , \  \ \ \ {\bar \r} \equiv  P\r P \ , \ \ \ \ \ \ \  {\bar
\r}  H\subseteq H_c \
.   $$ The reason why one finds  a conformal  model for an arbitrary $\r$ can
be understood in the
following way (for a related discussion see   \givkir\ST).
Consider the gauged  $[G\times H]/H$  WZW model  with $H$ isomorphic to  (a
subalgebra of) the Cartan
subalgebra of  $G$. Its action can be represented as the sum of the action of
the gauged $G/H$ WZW
model and the  gauged  action corresponding to $H$,
 $$I_H =  {1\ov 2 \pi}  \int d^2 z   (\del y_s +
\l_{st}B_t)  (\bd y_s+ \l_{st'}\B_{t'})\ , $$
 where $y_s \ ( s =1,...,r)  $   are the variables of the  WZW
theory for  $H$ and  $\l_{st}$ are  constants that parametrise the embedding of
$H$ into $G\times
H$. In the gauge $y_s=0$ this action is equivalent to (2.1)  if  $Q$ there is
given by (4.14$'$)
with  $\r_{st} = -4 \l^{-2}_{st} $.  Since the gauged  $[G\times H]/H$  WZW
theory  is conformal,
the model (2.1), (4.14$'$) should, of course,  also be conformal. The chiral
gauged WZW model
corresponds to the particular case of $\r=-2 I$.\foot{ This is a reflection of
the general equivalence
\ST\  of the $G/H$
chiral gauged WZW model  with an {\it abelian} $H$
to the  $[G\times H]/H$  gauged  WZW model with a special value of
the embedding parameter of the axial  subgroup  $H$ into the group $G\times
H$.}
The models (2.9), (4.14$'$) with  different  values of $\r_{rs}$  can be
generated  from the pure WZW
theory
 by the $O(2r,2r)$ duality transformations   corresponding to the
isometries along the Cartan algebra directions  \hussen\nap\kiri\givkir\ and
thus are also related by the duality. The second term in (2.9) in this case can
be put into the form
of an integrably marginal deformation \hussen\nap\kiri\givkir.

\def \brr {{\bar \r}}
The Hamiltonian corresponding to the solution (4.14$'$)  is  given by
(2.13),(2.18)
$$ \H  = \ha [I + \ha {\bar \r} (P+\brr)\inv \brr]  (J_a  J_b  + \J_a \J_b)
- \ha {\bar \r} (P+\brr)\inv (2P + \brr)   J_a\J_b  \ ,  $$
i.e.    contains the $J\J$-term.
The  central charge  (3.29)  for $K$ in (4.14$'$) is $\r$-independent and is
the same as for   the
 $[G\times H_c]/H_c$ coset  or simply $G$ affine-Sugawara  model, i.e. is given
by (4.5).

  One  may
look for other  solutions of (4.1)  representing  the symmetric   matrix  $K$
in the
`diagonal'
  form (as in
\halpo\halpoo) $$ K_{ab} = \sum_c p_c \Omega_{ac} \Omega_{bc} \ , \ \ \  \ \
\Omega^T \Omega = I  \ .
\eq{4.15} $$
Here $\Omega$ is an element of $SO(D_G)$.
 Then the basic equation (4.1) reduces to
$$ \sum_{k,l}
 \big[ p_m p_n {(p_k -p_l)^2 \ov p_k p_l}
- {(p_kp_l-1)^2 \ov p_k p_l} \big] {\hat f}_{mkl} {\hat f}_{nkl} =0 \ ,
\eq{4.16} $$
$$ {\hat f}_{mkl} \equiv  \Omega_{mm'}\Omega_{kk'} \Omega_{ll'} f_{m'k'l'} \ .
$$
The solutions of (4.16)  correspond to extrema of
the central charge `action' (3.30) which in the case of (4.15) is given by
$$ \C= D - {1\ov 32} \a' \sum_{m,k,l}( - p_m\inv p_k\inv p_l\inv - 3 p_m p_k
p_l\inv
+ 6 p_m + 6p_m\inv){\hat f}_{mkl}{\hat f}_{mkl}  + O(\a'^2) \ . \eq{4.17} $$
A special solution of (4.16)  is equivalent  to  (4.14$'$).
If $K^2=I$  we have  $p_n^2=1$
and   (4.16) becomes
$$
 \ (p_m p_n -1)\sum_{k,l}{(p_kp_l-1)}
 {\hat f}_{mkl} {\hat f}_{nkl} =0 \ . \eq{4.18} $$
%or, equivalently,
%$$
% \ (p_m - p_n )\sum_{k,l}{(p_k-p_l)}
% {\hat f}_{mkl} {\hat f}_{nkl} =0 \ . \eq{4.18'} $$
The non-zero  components
 of (4.18)   correspond to $m\not=n$ and $p_m =-p_n=1$;
then $\sum_{k,l}{(p_kp_l-1)} {\hat f}_{mkl} {\hat f}_{nkl} =0$.
The previously discussed chiral gauged WZW  solution   (3.27) is reproduced  as
a particular  case.

It is easy to show that no  additional  solutions  of (4.16) are found if
$G=SU(2)$
or $SL(2,R)$.  Here ${\hat f}_{mkl}=  f_{mkl}=\epsilon_{mkl}$ and
  we  get
from (4.16)  $$ p_1^2(p_2-p_3)^2=(p_2p_3-1)^2 \ ,
\ \ \ p_2^2(p_1-p_3)^2=(p_1p_3-1)^2\ , \ \ \ p_3^2(p_2-p_1)^2=(p_2p_1-1)^2 \ ,
\eq{4.19} $$
with the only non-trivial solution  (up to permutations and replacements of
$+1$ by $-1$) being
(4.14$'$), i.e. $ p_1=1\ , p_2= 1 , \ p_3= 1+ \r =$arbitrary. For $\r=-2$ we
have the  solution
(3.27).
For $G=SL(2,R)$  the resulting model (2.1),(3.1)--(3.4),(4.14$'$) is equivalent
to the `charged black string'  or  $[SL(2,R)\times R]/R$ gauged WZW model
\horne\
which,  in fact, is  conformally invariant for an arbitrary  value
of   one  free   parameter  (charge)  related to $\r$.

%%%%%%%%%%%%%%%%%%%%%%
\newsec{ Relation to Virasoro master equation}
%%%%%%%%%%%%%%%%%%%%%%%%%%%%%
An obvious question is how   eq.(4.1)   is   related  to the  Virasoro
master equation of refs.\hal\mor
$$ {L}^{ab} =   {  L}^{ac} { L}^{cb}  + {1\ov 2k} \bigl(    f^a_{\ cd}
f^{b'}_{\ cd'} { L}^{bb'}{
L}^{dd'} + f^{a'}_{\ cd} f^{b}_{\ cd'} { L}^{aa'}{ L}^{dd'} - f^a_{\ cd} f^b_{\
c'd'} { L}^{cc'}{
L}^{dd'}\bigr) \  .
 \eq{5.1}
 $$  We  have rescaled $L$ of \hal\halpo\ by $2k$.
  Using the same representation  for $L$ as
for $K$  in  (4.15)  one can put (5.1) into the form \halpo\halpoo
$$ \l_a (1- \l_a)\eta_{ab} = {1\ov 2k}  \sum_{c,d} \l_c(\l_a +
\l_b - \l_d) {\hat f}_{acd} {\hat f}_{bcd} \
, \ \ \ \ { L}^{ab} = \sum_c \l_c \Omega^{ac} \Omega^{bc} \ .  \eq{5.2} $$
In the  large  $k$ limit  the system  (5.2) reduces to  \halpoo
$$  L=L^2\ , \ \ \  \ \l_a (1-\l_a) =    0  \ , \eq{5.3} $$
$$ \sum_{c,d} \l_c(\l_a +
\l_b - \l_d) {\hat f}_{acd} {\hat f}_{bcd} =0  \ , \ \ \ a \not= b\
, \eq{5.4} $$
where $\hat f$ in (5.4) depends on the leading-order form of the `angular'
variables $\Omega$.

Since our equation (4.1) was derived in the leading order approximation in
$\a'=2/k$
it  should be compared  with (5.3),(5.4).
 While  all  $k\ra \infty$ solutions the master equation
correspond to $L$ being a projector our  equation (4.1) has also   solutions
with $K\not=K^2$.
 As  it is  clear from the structure of the Hamiltonian of our model
(2.13),(2.18),   a correspondence should be possible  only  in the special case
when our matrix $K$
satisfies $K^2=K=K\inv$,   i.e.  is  a projector on a subspace.
 Then we can identify  $L$ with $K$  (i.e. $p_a$ with $\l_a$). The equation
that follows from (4.1)
when $K^2=K$  is   (4.7). If one uses the representation (4.15) this equation
is the
same as (4.18) (now with  $p_a^2=p_a$). Though (4.18)  looks
  similar to (5.4) the two equations are not equivalent in general (but for
$SU(2)$ the solutions
are the same).

That (5.1) with $L^2=L$ is different from  (4.1) with $K^2=K$  (i.e. from
(4.7)) is easy to see also
without using the `diagonal' representation. In terms of  the same notation as
in
 (4.7$'$),(4.7$''$) ($f_{{\hat m} kl} = L^\perp_{mm'} f_{m'kl}, \  f_{\unn  kl}
 = L_{nn'}
f_{n'kl}$)  we can represent the  $LL^{\perp}$ and $L^{\perp}L^{\perp}$
projections  of (5.1)
(equivalent to (5.4))  in  the following way
 $$    f_{{\hat m }\unk l}f_{\unn\unk l}    - f_{{\hat m
}\unk\unl}f_{\unn\unk\unl}
= f_{{\hat m }\unk\hat l}f_{\unn\unk{\hat l} }  =0   \ ,  \eq{5.5} $$
$$  f_{{\hat m }\unk\unl}f_{{\hat   n}\unk\unl} =0 \ .
 \eq{5.6} $$
The system (5.5),(5.6)   is obviously different  from  (and is much less
restrictive than)
the system
(4.7$'$),(4.7$''$) that follows from  (4.1),(4.7).

The only obvious common solution of (4.1) and the $k\ra\infty$ limit  of (5.1)
is the coset one ($L=K=I-P$). The $k\ra\infty$ master equation,  i.e.
(5.3),(5.5),(5.6)  has  many
other solutions with $L=L^2$ \halpo\halpoo.
 At the same time, our equation (4.1) has also other solutions
with $K^2\not= K$, namely, (4.14$'$) and, in particular, the chiral gauged  one
(3.27).
The sets  of solutions of (4.1) and (5.3),(5.5),(5.6) thus intersect but do not
coincide.

 The  relation
  between  eq.(4.1) and the $k\ra \infty$ limit of the Virasoro master equation
(5.1)
is the following. The master equation describes only `irreducible' solutions
while (4.1) contains    `reducible' solutions corresponding to
{\it some} of the  solutions (cosets) of the master equation.
The `reducible' solutions  can be understood
as some `twisted' products of `irreducible' WZW models
 with `twisting' being due to  mixing of group
variables  and reducing
of the configuration space by integrating out some of the degrees of freedom
(group variables that parametrise the $2d$  gauge field).

 For example, it is clear from (2.8)
that before one integrates out the  $2d$  gauge field the action of the chiral
gauged WZW model
is just a sum of the three WZW actions so  that the corresponding stress tensor
is given by   the
three independent affine-Sugawara terms each of which is of course a solution
of the master equation.
At the same time, the Hamiltonian (2.16) one finds upon elimination of the
gauge field is not of  the
standard coset model type. The reason why it still corresponds to a conformal
theory
(i.e. represents  a solution of (4.1)) is that the currents that appear in
(2.13),(2.16)
are {\it not chiral} (as it is assumed in the affine--Virasoro construction).
In fact, if the
$J\J$-term in (2.9) is {\it not} treated just as a perturbation of the WZW
theory the currents are no
longer (anti)holomorphic on the equations of motion.
 As was mentioned in \ST, it should be possible to define the new {\it
(anti)holomorphic} currents (combinations of $J$,  $\J$, $J_H$ and $\J_H$) in
terms of  which the
Hamiltonian will take again  its   standard Sugawara-like form.   Similar
remark applies to the
general case of the models (4.14$'$)  since they are equivalent to the
$[G\times H_c]/H_c$ coset
 models.\foot{Since  the basis of  the affine-Virasoro construction is the
current algebra,  the
starting point in \hally\  is the classical Hamiltonian  which does not contain
 $J\J$-terms. This is
natural since  {\it if } the currents   are (anti)holomorphic then  such a
structure  is implied by
conformal invariance. At the same time, the currents  that appear in  the
Hamiltonian of our model
(2.13)  are not, in general,   (anti)holomorphic on shell (since the equations
of motion that
follow from  (2.1),(2.9) are {\it different} from the standard equations of the
WZW  model).
As a result, we got  conformal  solutions  (4.14$'$) (with $K\not=K\inv$)  for
which  there {\it is}
a  $J\J$-term in the Hamiltonian.}

In general, our  model (2.1),(4.1)  should not be expected to describe
`reducible' solutions
corresponding to `irreducible' solutions of the master equation other than
cosets.  In fact,
since $K$ is the constant part of  the metric  (3.18)  the derivation of (4.1)
from the
$\bar \b$-function equations (3.8) formally applies only when $K$ is
non-degenerate.
The case of the gauged WZW model when
 $K=K^2$, i.e.  is singular, is a special one; it can still be  treated in a
consistent
way   because  of the explicit gauge invariance of the action (2.1) when
$K=I-P$.  It is  clear,
however,  that  other solutions of the master equation with $K=L=L^2$ should
fall outside  of our
class of models (2.1).  In fact, it  is known that
 an  action  \hally\ that   reproduces  the  off conformal point extension of
the (large $k$)
Hamiltonian of  a generic {\it  non-coset}  affine--Virasoro construction  is
{\it not}  Lorentz
invariant  \hally\halpe.  When $L$ is subject to the master equation the
action of \hally\ is Lorentz invariant provided one does not ignore its
dependence on extra  degrees
of freedom (Lagrange multipliers) $v_{zz}$ and  ${\bar v}_{{\bar z }{\bar z}}$.

 To summarise,  the  solutions of equation (4.1)  are  only the `reducible'
counterparts of  such
 solutions   of the master equation  (cosets) that can be described by  field
theories that are
manifestly Lorentz invariant off the conformal point.
A natural problem    then  is to find an  analog of (4.1)
 which will  contain  more general solutions of the master equation  (as well
as their
`reducible' counterparts) by starting  from a non-Lorentz-invariant
field-theoretic model and
imposing the conditions of conformal/Lorentz invariance at the quantum level.
An
existence of such generalised \sms  that are not
Lorentz invariant at the classical level but  become invariant  at the
conformal point
was conjectured in \tsdu\ (where the \sms with doubled number of coordinates
were introduced in
order to make the target space  duality symmetry  manifest at  the string world
sheet  action
level).

One possible starting point is the group space  action of \hally\ (in the gauge
$v=\bar v =0$).
We would like, however, to suggest  what seems to be a  natural  alternative
%(but possibly equivalent)
 approach which is   based on  doubling of the number of  group space variables
(but not of the
degrees of freedom). The idea is that  in trying to construct an  off shell
extension, one may
represent the chiral currents $J, \ \J $ of the affine-Virasoro construction
either in
terms of  one group field $g(z,\bar z)$  with the standard  WZW equations of
motion  or a pair of
 fields
 $g_-(z,\bar z)$, $g_+(z,\bar z)$  which  become chiral on shell.
 There exists a  reformulation  of the WZW model in which one
replaces the  field $g(z,\bar z)$ by the two `chiral' fields $g_-(z,\bar z)$
and $g_+(z,\bar z)$
described by   the Floreanini-Jackiw type \floja\  WZW Lagrangians \son\
(cf. (2.2))\foot{These actions can be also  obtained from
the manifestly Lorentz invariant  actions in the Siegel's
approach to chiral scalars \sigel\ by gauge-fixing the Lagrange multiplier
\chirwzw.}
$$ I_\pm (g_\pm) =
   \pm {1\over 8  \pi }
\int d^2 z  \Tr (\del_1 g^{-1}_\pm
\del_\mp g_\pm )  +  {1\over  12 \pi   } \int d^3 z \Tr ( g^{-1}_\pm dg_\pm)^3
  $$
$$ = {1 \over  8 \pi  } \int d^2 z  \bigl[  { G}_{0 mn }  (x_\pm) ( \pm
\dx^m_\pm  x'^n_\pm  -
x'^m_\pm x'^n_\pm)  + B_{0 mn} (x_\pm) \dx^m_\pm x'^n_\pm  \bigr]  \ ,
\eq{5.7} $$
or
$$ I_\pm (g_\pm) =    I'_\pm (g_\pm)  -
{1 \over  8 \pi  } \int d^2 z    { G}_{0 mn }  (x_\pm)
x'^m_\pm x'^n_\pm\ , \   $$ $$  I'_\pm (g_\pm)\equiv
 {1 \over  8 \pi  } \int d^2 z    (\pm { G}_{0 mn } + B_{0mn}) (x_\pm)
\dx^m_\pm
x'^n_\pm   \ .    \eq{5.8} $$
These models  are Lorentz invariant only on the equations of motion.
In the absence of interaction between
 $g_-(z,\bar z)$ and $g_+(z,\bar z)$
 it is possible to integrate out the
`ratio' of $g_-$ and $ g_+$  explicitly,  ending up with the standard WZW
action for $g=g_-g_+$
\sto\tsdu.   Let us  consider the following analog  of (2.1)
 $$ I_L (g_\pm, A_\pm) = I_+(g_+)  + I_-(g_-) $$ $$  +{1\over 2\pi }
 \int d^2 z \Tr \bigl[ - A_- (L-I) J_+    + J_- (L-I) A_+
-    A_- (L-I) A_-  -  A_+ (L-I) A_+  \bigr] \ , \eq{5.9}  $$
$$  J_+ =g\inv_+\del_1 g_+ = i T_a E^a_m x'^m \ , \ \ \ \
\ \ J_-=\del_1 g_- g\inv_- = i T_a \E^a_m x'^m \  , \eq{5.10} $$
where $L^{ab}$ is a constant symmetrix matrix. Integration over $A_\pm$ gives
(cf. (2.9))\foot{A
model of this type was considered in  \tsdu.  Similar actions in the
Siegel's formulation \sigel\ for chiral scalars appeared in \gat. The latter
approach is
manifestly Lorentz invariant but one has to deal with extra (Lagrange
multiplier) degrees of freedom
and preserve \hull\ the corresponding gauge symmetries.  Such  actions  with
Lagrange multipliers may
be  related to the action in \hally.}
 $$ I_L (g_\pm) = I_L(g_+) + I_L(g_-) = I_+(g_+)  + I_-(g_-)  + {1\over  8 \pi
}
 \int d^2 z \  \Tr \bigl[
     J_- (L-I)J_-       +       J_+ (L-I)J_+  \bigr]  \eq{5.11}  $$
$$=  I'_+(g_+)  + I'_-(g_-)  + {1\over  8 \pi }
 \int d^2 z \  \Tr \bigl(
     J_- LJ_-       +       J_+ LJ_+  \bigr) \ , \eq{5.12} $$
or
$$ I_L = {1\over  8 \pi }\int d^2 z  \bigl[  ( { G}_{0 mn } + B_{0mn}) (x_+)
\dx^m_+ x'^n_+  -    (
{ G}_{0 mn } - B_{0mn}) (x_-)   \dx^m_-  x'^n_- $$ $$
- L^{ab} E_{am}E_{bn} x'^m_+x'^n_+  - L^{ab} \E_{am}\E_{bn} x'^m_-x'^n_- \bigr]
\ . \eq{5.13} $$
Since the actions $I'_\pm$ are linear in time derivatives,  the
corresponding Hamiltonian is proportional to the third term in (5.12). In fact,
 one can consider
each of $x^m_+$ and $x^m_-$ as phase space coordinates, i.e. as  a mixture of
`true' coordinates
and momenta \faj.
If the Lagrangian of a mechanical system  is $\ L = a_{i}(q) {\dot q}^i  -
V(q), \ $ then the
Hamiltonian  is just  $\H= V$ and
the Poisson bracket  of functions on the phase space is
$ \{ X_1(q), X_2 (q)\}= F^{-1 ij} \del_i X_1 \del_j X_2   , \ \
F_{ij} \equiv \del_i a_j -  \del_j a_i \ $ \witt\faj.
It is possible to check that if  one starts with
the action (5.11) or, equivalently, with  $I_\pm (g_\pm)$ or $I'_\pm (g_\pm)$,
then the brackets one gets are
such that  the currents (5.10) form the  standard affine algebras \son.
The Hamiltonian for  the pure `kinetic' action $I'_\pm$
is zero,  for the chiral  WZW action $I_\pm $
is given  by the `potential' $\del_1 g\inv_\pm \del_1 g_\pm$ term or
$ \H_\pm  =    \ha \eta^{ab} J_{\pm a} J_{\pm b} , $ while for (5.12) is
obviously
$$\H= \ha L^{ab} (J_{+ a} J_{+ b}
+ J_{- a} J_{- b})\ . \eq{5.14} $$
We conclude that the action (5.11)  may be considered as an alternative
Lagrangian
realisation of the  Hamiltonian of the  affine-Virasoro construction
\hal\mor.\foot{A
suggestion that a  different `Thirring-like'  action  for  two interacting  WZW
fields $g_L$ and $g_R$
may correspond to the solutions of the master equation was made in \solo.
However, we believe the
approach of \solo\ is not consistent since the fields $g_L$ and $g_R$ where
assumed to be
`non-chiral' having   manifestly Lorentz invariant action  (i.e. describing
doubled number of degrees
of freedom)  while the chirality constraint was imposed later `by hands'.
This does not seem to go
beyond  a trivial rephrasing of the original affine--Virasoro construction.}
 While in \hally\ the current algebra was represented in terms of one set of
group coordinates
of the standard WZW action, we got a simpler action by using two `chiral' sets
$x^m_+$ and $x^m_-$.
In the case when $L$ satisfies (the large $k$ form (5.3) of) the master
equation
the  Hamiltonian system has invariance implied by  the conjugation invariance
$L'= I-L$
\hally;  as in \hally\  the corresponding constraints  can be added  (with
Lagrange multipliers
$v, \bar v$)   to the action  (5.12) by replacing $L$ by $  L +v(I-L)$ in the
`$+$' part and $L$ by $
L +{\bar v}(I-L)$ in the `$-$' part.
The classical equations that follow from (5.12) can be represented as two
separate equations for the
currents $J_+$ and $J_-$ (symbolically, ${\dot J}_+ = f L J_+J_+ + L J'_+$,
etc) and are
nothing but the Hamiltonian equations $\dot J_\pm = \{ \H, J_\pm \} $
corresponding to (5.13)
if $J_\pm$  there form commuting affine algebras.

 An advantage of the action (5.11)
(5.9) is that it naturally incorporates the  case  of the coset solution
$L=I-P_H$ where $P_H$ is a
projector on a subalgebra $H$. If we assume that $A,\A$ take values in $H$
eq.(5.9) becomes
 $$ I_{G/H} (g_\pm, A_\pm) = I_+(g_+)  + I_-(g_-)   +{1\over 2\pi }
 \int d^2 z \Tr \bigl( - A_- J_+    + J_- A_+
-    A_-  A_-  -  A_+  A_+  \bigr) \ . \eq{5.15}  $$
This is  just the sum of the  analogs of the gauged WZW action in the `chiral'
case.
In fact, consider, e.g.,
$$ I_{G/H} (g_+, A_-) = I_+(g_+)   + {1\over  2 \pi }
 \int d^2 z \  \Tr \bigl( - A_- J_+
-  A_- A_-         \bigr) =  I_+(hg_+ )  - I(h)   \ ,   \eq{5.16}  $$
where  $A_-= \del_- h h\inv$ ($h$ is from $H$) and $I(h)$ is the usual WZW
action (see also \son).
 The action (5.16) is invariant under
$ g_+ \ra  fg_+  , \ \ h\ra hf\inv , \ f=f(x_-)$.\foot{To have a non-chiral
gauge invariance one
needs to add an extra term $A_-A_+$. This  corresponds to a `vector'
regularisation  scheme. In
the  present setting the `left-right decoupled' scheme seems more natural. If
$A_-A_+$-term
is added to (5.9) the action (5.11) and the Hamiltonian (5.14) become more
complicated and, in
particular,  contain $J_+J_-$-term. The absence of the $A_-A_+$-term is the
reason
why by  integrating out the `ratio' of $g_+$ and $g_-$ one  gets  not a
standard gauged WZW
action but a chiral gauged WZW action for
$g=g_+g_-$.}

The action (5.12) is not, in general, Lorentz invariant  at the classical level
(even  on the equations of motion).
The coset case  $L=I-P_H$ (including $L=I$)  is special since here the action
(5.12) {\it is}  Lorentz
invariant on the mass shell. This is not surprising since in the coset  case
 one can integrate out the `ratio' of $g_+$ and $g_-$  explicitly,   getting a
local,
manifestly   Lorentz invariant action for $g=g_+g_-$.
In fact,  a combination of (5.16) with a similar action for $g_-, A_+$ leads to
a (chiral) gauged WZW  action for $g=g_+g_-$.  Since  \tsdu\ \foot {To prove
(5.17)
one changes the variables  $g_+= g^{1/2} f , \ g_-= f\inv g^{1/2}$.
Then the sum of the actions in
(5.17) becomes equal to $I(g)$ plus an aditional term  of the structure  $ \int
(B +  T(g))^2 , \
B\equiv \del_1 f f\inv$.  The integral over $f$  gives a trivial  contribution
since one can
replace $f$-integral by  the integral over   $B$-variable (the Jacobian is
trivial
since $\del_1\inv = \theta (z_1-z_1')$).}
$$ I_+(g_+) + I_-(g_-) \ \ra  \  I(g ) \ , \ \ \ \ \ \ g= g_+g_-\ \ , \eq{5.17}
$$
we get (cf. (2.8))
$$ I_+(g_+, A_-) + I_-(g_-, A_+) \ \ra  \  I_{chir} (g, A_+, A_-) \ . \eq{5.18}
$$
It may be possible to do a similar integration for some other special values of
$L$
returning back to the Lorentz invariant model (2.1) for $g=g_-g_+$.
One may expect  that  the condition of (one-loop)
conformal invariance  of the non-Lorentz-invariant models (5.9),(5.12)  can be
put  into
correspondence with  the master equation (5.3),(5.4).  Then  (5.9) would
describe `reducible' counterparts of  both the coset and non-coset
`irreducible' solutions of the
master equation.

In conclusion, let us mention that eqs.(3.2)--(3.3) give  the  universal
expressions for the
basic target space fields for the models of the class (2.1) making  it possible
 to study the corresponding geometries in a systematic way. Another open
problem is to find if there
are  other solutions of the conformal invariance equations (4.1),(4.16)  in
addition to
gauged WZW, non-abelian chiral gauged WZW and (4.14$'$).

\bigskip
{\bf Acknowledgements}

\noindent
I am grateful to  M. Halpern  for  very useful  discussions and criticism and
 would like to thank K. Sfetsos for comments.
I acknowledge  also  the  support of SERC.

 \vfill\eject
\listrefs

\end